\documentclass[useAMS,usenatbib]{mn2e}
\usepackage{graphicx,amsmath,color,amssymb}
\usepackage[pdftitle={},pdfauthor={Simeon Bird}]{hyperref}
\usepackage[all]{hypcap}
\usepackage{subfigure}

\newlength{\narrowFigurewidth}
\newlength{\Figurewidth}
\newlength{\wideFigurewidth}
\newcommand{\etal}
  {et al.}

\newcommand{\Lya}{Lyman-$\alpha\;$}
\newcommand{\Msun}{\, h^{-1} M_\odot}

\newcommand{\NHunit}{cm$^{-2}$}

\newcommand{\Mpch}{\, h^{-1} \mathrm{Mpc}}
\newcommand{\kpch}{\, h^{-1}\mathrm{kpc}}

\newcommand{\kms}{km~s$^{-1}$}

\newcommand{\cloudy}{{\small CLOUDY\,}}
\newcommand{\arepo}{{\small AREPO\,}}

\newcommand{\civ}{C$\mathrm{IV}$}
\newcommand{\cii}{C$\mathrm{II}$}
\newcommand{\SiII}{Si$\mathrm{II}$}
\newcommand{\SiIV}{Si$\mathrm{IV}$}

\title[Simulating the Carbon Footprint of Galactic Halos]{Simulating the Carbon Footprint of Galactic Halos}
\author[S. Bird \etal]{Simeon Bird$^{1,2}$\thanks{E-mail: sbird4@jhu.edu}, Kate H.~R.~Rubin$^{3}$, Joshua Suresh$^{3}$, Lars Hernquist$^{3}$
\vspace{4mm}\\
$^1$Carnegie Mellon University, 5000 Forbes Avenue, Pittsburgh, PA, 15206, USA\\
$^2$Johns Hopkins University, Baltimore, MD 21218, USA\\
$^3$Harvard-Smithsonian Center for Astrophysics, 60 Garden Street, Cambridge, MA 02138, USA
}

\begin{document}

\maketitle

\begin{abstract}
    We compare simulations, including the Illustris simulations, to observations of \civ~and \cii~absorption at $z=2-4$. 
    These are the \civ~column density distribution function in the column density range $10^{12} - 10^{15}$ \NHunit, the \civ~equivalent width distribution at $0.1 - 2$~\AA, and 
    the covering fractions and equivalent widths of \civ~$1548$~\AA~and~\cii~$1337$~\AA~around DLAs. In the context of the feedback models we investigate, all \civ~observations 
    favour the use of more energetic wind models, which are better able to enrich the gas surrounding halos. We propose two ways 
    to achieve this; an increased wind velocity and an increase in wind thermal energy. However, even our most energetic wind models do not produce enough absorbers 
    with \civ~equivalent width $> 0.6$~\AA, which in our simulations are associated with the most massive haloes.
    All simulations are in reasonable agreement with the \cii~covering fraction and equivalent widths around Damped Lyman-$\alpha$ absorbers, although there is 
    a moderate deficit in one bin $10 - 100$ kpc from the DLA. Finally, we show that the \civ~in our simulations is predominantly photoionized.
\end{abstract}

\begin{keywords}
intergalactic medium -- galaxies: formation
\end{keywords}

\section{Introduction}

Understanding the nature of galactic feedback, the manner in which luminous objects such as stars and quasars affect the gas surrounding them,
is one of the most significant open problems in galaxy formation. Neither individual supernovae nor black hole accretion discs are resolved 
in current cosmological simulations, and so they are included through approximate effective models for stellar and Active Galactic Nuclei (AGN) feedback.
These models are tuned to produce realistic low-redshift galaxies \citep[e.g.][]{Schaye:2010, Dave:2011, Puchwein:2013, Vogelsberger:2013, Schaye:2015}.
Observations apart from those the models were adjusted to match are also reproduced. For example, the stellar properties of high redshift 
galaxies \citep{Torrey:2013}, the neutral hydrogen column densities in Damped \Lya Systems (DLAs) \citep{Bird:2014} or 
the abundances of Lyman Limit Systems around quasars \citep{Rahmati:2015}.
Further refining the sub-resolution modeling requires observational probes able to constrain independent parameters and thus inform different elements of the model.

In this paper we examine results from the Illustris Project \citep{Vogelsberger:2014b, Vogelsberger:2014, Genel:2014, Nelson:2015}, as well as smaller simulations
with a different supernova wind energy per unit mass. We compare our simulations to observations of absorption lines from singly and triply-ionized carbon.
\cii~traces relatively dense, low temperature gas closely associated with galactic haloes. 
\civ~traces enriched gas at a density more typical of the intergalactic medium, and thus provides complementary information 
to measurements characterizing on stars or neutral hydrogen \citep{Haehnelt:1996}.
Examining both ions helps us to distinguish variations in ionization fraction from variations in enrichment.
\cite{Suresh:2015} showed that this parameter affects the distribution of 
circumgalactic metals, and that an increased energy can help improve agreement with measurements of the \civ~column density 
around $z \sim 2.5$ galaxies \citep{Turner:2014}. Here we show that this parameter has further observational consequences. 

We first compare our simulations to the line density and column density function of weak (column density $< 10^{15}$ \NHunit) \civ~absorbers \citep{DOdorico:2010}.
\footnote{These absorbers are sometimes taken to trace the metal content of the Intergalactic Medium (IGM), but \cite{Oppenheimer:2006} showed 
the two are poorly correlated.} Comparisons of this type have already been used to constrain feedback models. 
Indeed, \cite{Oppenheimer:2006, Oppenheimer:2008} (henceforth OD06, OD08) used data from \cite{Songaila:2001, Boksenberg:2003, Songaila:2005} and \cite{Scannapieco:2006}
to help motivate the class of supernova feedback models on which the Illustris Project is based. \cite{Tescari:2011} expanded the analysis to include the 
updated data of \cite{DOdorico:2010} and to constrain a broader class of feedback models, including AGN feedback.
Most recently, \cite{Rahmati:2015a} compared the results of the EAGLE simulation to observations of \civ, and other high ionization systems.
However, our study will be the first to use \civ~to test the feedback model in the Illustris simulation. We focus on comparisons to the dataset 
presented in \cite{DOdorico:2010} as their survey covers a combined path length of $\Delta X = 86$, the largest \civ~survey currently available at high resolution. Furthermore, using a single recent measurement 
of the \civ~column density function allows us to generate simulated spectra which match as closely as possible the properties of the 
observations.

In addition, we will perform a comparison to other observed properties of \civ. We use the line density of strong \civ~absorbers 
(equivalent width $0.3 - 1.2$~\AA) from \cite{Cooksey:2013}, measured using a large spectral sample of absorbers from the Sloan Digital Sky Survey \citep[SDSS;][]{York:2000}. 
These stronger absorbers allow us to extend the range of our comparison to gas at a density and enrichment found only around large haloes. 
To check our simulations against smaller haloes traced by high redshift galaxies too faint to be detected in imaging surveys, we compare to the carbon absorption 
around DLAs as measured by \cite{Rubin:2014} using close quasar pairs. 

The literature contains various theoretical studies examining how different feedback models influence the properties, ionization state and 
enrichment of the $z=2$ IGM and circumgalactic medium (CGM). For example, \cite{Cen:2011} and \cite{Barai:2013} showed that allowing wind strength to vary as a function 
of galactic radius affects the gas and metallicity profiles. \cite{Ford:2013, Ford:2014} found that the CGM at $z=0.5$ is enriched by 
cool outflows which then ultimately re-accrete onto the galaxy. By contrast, \cite{Shen:2012, Shen:2013} found that at $z=3$, outflows free-stream out of the
potential well of a simulated high resolution galactic halo. \cite{Suresh:2015} showed that these differences arise not only due to the differing redshifts, 
but also owing to their wind models having different effective energy per unit mass. As the theoretical properties of the CGM have already been extensively examined, 
we choose to focus here on a comparison to specific observations.

We use observations in the redshift range $z=2-4$, in part because \cite{Suresh:2015} showed that this 
is the epoch during which the circumgalactic metallicity is most sensitive to the parameters of the supernova model. 
The gas metallicity and \civ~abundance at lower redshifts are controlled by the AGN feedback model and at higher redshifts by the details of reionization. 
Furthermore, the statistical power of the observational surveys is largest at $z=2-4$.
The evolution of the IGM from $z=2-0$ has been examined by \cite{Dave:2010}, while \cite{Oppenheimer:2012} compared their simulations to low-redshift ($z = 0-0.5$) 
metal-line data \citep{Danforth:2008, Cooksey:2010}, and more recently \civ~surveys have been carried out at $z<2$ by \citep{Burchett:2013, Burchett:2015}. 
Future work could use these surveys to constrain AGN feedback models.
Constraints on the \civ~column density distribution has also been obtained at $z > 5$ by \cite{Pettini:2003, Becker:2009, RyanWeber:2009, Simcoe:2011}. 
The implications of our simulations for these observations was examined in \cite{Keating:2016}.

In Section \ref{sec:methods}, we describe our simulations and our methods for generating mock observations. Section \ref{sec:absorbers}
discusses properties of our simulated absorbers, including their host haloes and the extent to which their ionization state is dominated 
by photo-ionization. Section \ref{sec:observations} presents a comparison to the observations, and we draw conclusions in 
Section \ref{sec:conclusions}. Appendix \ref{sec:resolution} shows the convergence of our results with numerical resolution, and Appendix \ref{sec:tables} 
reproduces in tabular form the results shown in some of our figures.

\section{Methods}
\label{sec:methods}

In this section we explain our methods, including a brief overview of the simulations (section~\ref{sec:simulations}) and feedback models 
(section~\ref{sec:feedback}). We then describe briefly how we compute our synthetic spectra (section \ref{sec:spectra}).
Analysis codes specific to this paper are available at \url{https://github.com/sbird/civ_kinematics}.

\subsection{Simulations}
\label{sec:simulations}

The simulations analysed in this paper are summarized in Table \ref{tab:simulations}. They comprise the Illustris Project and 
a few variants on its feedback model, originally described in \cite{Bird:2014}.
All simulations were run using the moving mesh code \arepo\ \citep{Springel:2010}, which combines the TreePM method 
for gravitational interactions with a moving mesh hydrodynamic solver.
Each grid cell on the moving mesh is sized to contain a roughly fixed amount of mass, and to move approximately following
the bulk motion of the fluid locally. Small-scale mixing is included by allowing gas and metals to advect between grid cells.

The largest simulation we employ, hereafter ``Illustris'', is a box with a comoving linear size $75 \Mpch$ and 
a comoving gravitational softening length of $\sim 1 \kpch$. Illustris initially contains $1820^3$ dark matter (DM) particles and 
$1820^3$ gas elements. It is described in more detail in \cite{Vogelsberger:2014b, Vogelsberger:2014} and \cite{Genel:2014}.
We have also run several smaller simulations, each with a comoving linear size of $25 \Mpch$, initially 
containing $2\times 512^3$ resolution elements and having a comoving gravitational softening length of $\sim 1 \kpch$. 
They thus have a similar mass resolution to the Illustris simulation but a lesser volume. 
Illustris has both stellar and AGN feedback, described in detail in \cite{Vogelsberger:2013}, and summarized in Section \ref{sec:feedback}.
Our smaller simulations vary parameters of the stellar feedback model, as described in Section \ref{sec:feedback}.

\begin{table}
\begin{center}
\begin{tabular}{|l|c|c|c|l|}
\hline
Name & Box Size& AGN & $\kappa_\mathrm{w}$ & Notes \\
	& (Mpc / h) & 	&	(Eq. \ref{eq:kappa})		& \\
\hline 
ILLUS 75    &  75       & Yes & $3.7$ & Illustris simulation\\  
ILLUS 25    &  25       & Yes &  $3.7$ & \\ 
WARM  25 &  25       & No &  $3.7$ & Thermal winds \\ 
FAST  25   &  25       & Yes &  $5.5$ & \\ 
\hline
\end{tabular}
\end{center} 
\caption{Table of simulation parameters. The AGN column shows whether AGN feedback is enabled. $\kappa_\mathrm{w}$ sets the wind velocity.
$50\%$ of the wind energy in the WARM simulation is deposited as thermal energy, as explained in the text.
}
\label{tab:simulations}
\end{table}

\subsection{Feedback Models}
\label{sec:feedback}

\begin{figure}
\includegraphics[width=0.45\textwidth]{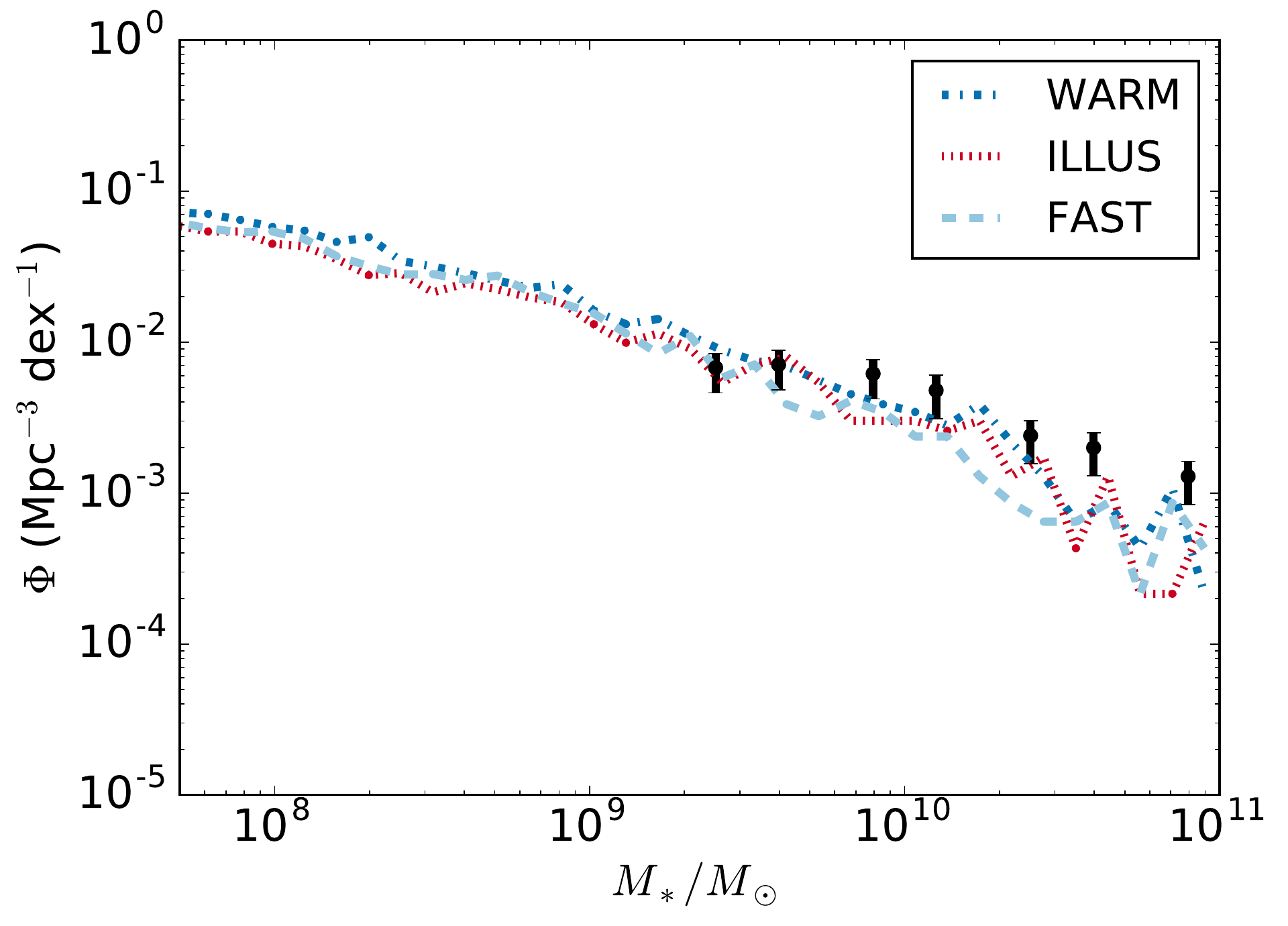}
\caption{The galactic stellar mass function at $z=2$ for our three feedback models, WARM, FAST and Illustris, compared to the galactic stellar 
mass function observed at $z=2-2.5$ by \protect\cite{Mortlock:2011}. While there are no direct constraints for faint objects at this redshift, all our models
agree with the measurements where they overlap. Furthermore each model has a very similar stellar mass function.
}
\label{fig:gsmf}
\end{figure}

As described in detail in~\cite{Vogelsberger:2013}, Illustris includes phenomenological models for feedback which aim to capture the unresolved influence of supernovae
and AGN on their environment. The parameters of these models have been adjusted to 
approximately reproduce the galactic stellar mass function and star formation rates at $z=0$, by suppressing star formation relative
to pure gravitational collapse. Mass and energy are returned from star-forming cells to nearby gas cells, adding kinetic and thermal 
energy and suppressing star formation. This is implemented by star-forming cells stochastically creating wind particles with an energy roughly corresponding to the expected available
supernova energy per stellar mass. Wind particles are collisionless. To prevent the wind energy from being quickly dissipated through 
radiative losses, they are decoupled from hydrodynamic evolution until they reach lower density regions or hit a time limit. 
At this time they are removed and their energy, momentum, mass and metals are added to the gas cells at their current location.
As the hydrodynamic decoupling models the effect of unresolved density fluctuations on the radiative cooling of supernova shock fronts, 
the parameters of the feedback model should be viewed as being a function of the decoupling time, and must be re-tuned when this changes.

In order to adequately suppress star formation in faint objects, the mass returned from 
star-forming regions scales with the local velocity dispersion of the dark matter (DM).
The total energy of each wind particle, $\mathrm{egy}_\mathrm{w}$, is given by 
\begin{equation}
 \mathrm{egy}_\mathrm{w} = \frac{1}{2} \eta_\mathrm{w} v_\mathrm{w}^2\,,
 \label{eq:energy}
\end{equation}
where $\eta_\mathrm{w}$ is the wind mass loading and $v_\mathrm{w}$ is the 
wind velocity. $\mathrm{egy}_\mathrm{w}$ is constant within all our simulations. $v_\mathrm{w}$ scales with the local DM velocity 
dispersion, $\sigma^\mathrm{1D}_\mathrm{DM}$, which correlates with the maximum DM circular 
velocity of the host halo \citep{Oppenheimer:2008}. We define
\begin{equation}
 v_\mathrm{w} = \kappa_\mathrm{w} \sigma^\mathrm{1D}_\mathrm{DM}\,.
 \label{eq:kappa}
\end{equation}
The Illustris wind model thus yields large mass loadings in small haloes (as $\mathrm{egy}_\mathrm{w}$ is constant), which 
allows it to roughly match the galaxy stellar mass function at $z=0$ \citep{Okamoto:2010, Puchwein:2013}.
The specific parameters chosen yield in Illustris wind ejecta which are frequently not completely unbound from their originating halo.
Furthermore, the high mass loading leads to a relatively cold CGM \citep{Suresh:2015}.

The metallicity of the supernova winds is taken to be $0.4$ times the metallicity of the star-forming gas 
launching the wind, following \cite{Vogelsberger:2013}.
We ran a simulation where the supernova wind metallicity was $0.7$ times the gas metallicity and checked that this parameter had a negligible effect on
to each set of observables we consider. \cite{Vogelsberger:2013} and \cite{Suresh:2015} performed similar checks with the same result. This may indicate 
that the CGM is not enriched by the wind particles themselves, but by outflows accelerated by energy transferred from the wind particles.
Note that zero-metallicity winds can affect the dynamical evolution of the CGM over time by reducing the impact of metal cooling \citep{Suresh:2015}.
Metal cooling is included as outlined in \cite{Vogelsberger:2013}.

As shown in Table \ref{tab:simulations}, we consider two variants of this model: faster winds (FAST) and hotter winds (WARM).
Both of these have an increased wind energy per unit mass, $\mathrm{egy}_\mathrm{w} / \eta_\mathrm{w}$, while 
the total wind energy, $\mathrm{egy}_\mathrm{w}$, is kept fixed. The WARM model offers a way to include thermal energy without altering 
the stellar mass function. To accomplish this, the wind velocity is held fixed and half the total wind energy is deposited in thermal 
energy when the wind re-couples, following \cite{Marinacci:2013}. Thus Eq. \ref{eq:energy} becomes
\begin{equation}
\mathrm{egy}_\mathrm{w} = \frac{1}{2} \eta_\mathrm{w} v_\mathrm{w}^2 + \mathrm{egy}_\mathrm{therm}\,.
\label{eq:thermenergy}
\end{equation}
Since $\mathrm{egy}_\mathrm{w}$ and $v_\mathrm{w}$ are held fixed, increasing $\mathrm{egy}_\mathrm{therm}$ decreases the mass loading, $\eta_\mathrm{w}$
and hence increases $\mathrm{egy}_\mathrm{w} / \eta_\mathrm{w}$. Physically, thermal energy models shocking that may occur in the wind during the phase when it is modelled by a decoupled particle. 
\cite{Suresh:2015} showed that this wind model does not directly produce hotter gas near the virial radii of our dark matter haloes, a range $50-200 \kpch$ from the galaxy.
This is because outflows generally shock before reaching this point, even in the Illustris model. However, \cite{Suresh:2015} also showed that the 
higher energy outflows present in WARM can more easily escape the enclosing halo and enrich the CGM than in Illustris.

The star formation rate in a resolved halo is largely independent of particle mass resolution in all our simulations.
Adding extra thermal energy to the winds does not induce a resolution dependence, because thermal wind energy, like kinetic wind energy, is initially hydrodynamically 
decoupled from the gas. The feedback model governs how high density material couples to the rest of the simulation, and includes physical processes smoothed to the new density scale 
introduced by re-coupling. The numerical parameters of the feedback model are implicitly defined as a function of this scale, rather than the mass resolution of the simulation.
Thus changing the mass resolution of the simulation does not require recomputing these parameters to achieve the same star formation rate.

Our second parameter variation considers faster winds (FAST). Here we fix the wind energy, $\mathrm{egy}_\mathrm{w}$, as before, but increase 
$\kappa_\mathrm{w}$ to $5.5$, and hence increase the wind velocity $v_\mathrm{w}$ by $50\%$. Like warm winds, this enhances the enrichment of the circumgalactic medium \citep{Suresh:2015}. 
As shown in \cite{Vogelsberger:2013}, increasing the wind speed can reduce star formation at low redshifts, as faster outflows are 
more rarely re-accreted onto haloes. To avoid this, we increase the wind speed by a moderate factor of $1.5$, which enriches the CGM but, 
as shown by Figure~\ref{fig:gsmf}, does not affect the galaxy stellar mass function at $z\geq 2$.

Figure \ref{fig:gsmf} shows the galaxy stellar mass function for all three of our feedback models. Each feedback model yields similar results. The reduced wind mass loading 
in the WARM winds model weakens the strength of the feedback and slightly increases the stellar mass, while the faster winds in the FAST model have 
the opposite effect, especially for larger halos.
All three models agree well with the limited observational data available at this redshift. We have also checked that the stellar metallicities are similar for each model. 
In \cite{Bird:2014} we showed that all three models produce similar abundances and metallicities for Damped Lyman-$\alpha$ systems, strong neutral hydrogen absorbers. Thus none 
of our feedback models change the stellar or neutral gas properties of the simulation.
However, as explained in Appendix B2 of~\cite{Suresh:2015}, the default wind model of \cite{Vogelsberger:2013} launches winds with a characteristic energy
close to the virial temperature of the host halo. Both our parameter variations increase the energy per unit mass available to the outflows by 
approximately a factor of two, and thus launch winds exceeding the potential energy of their host halo that can enrich the circumgalactic 
medium to greater distances.

AGN feedback was disabled in the WARM winds model, to test the extent to which \civ~abundance at $z=2$ is sensitive to the AGN. As implemented in Illustris, AGN feedback suppresses star formation
in the most massive haloes by periodically releasing thermal energy from the black hole into the gas cells surrounding it. The amount of energy released
is chosen to be sufficiently large to avoid being dissipated through radiative cooling in the dense gas immediately surrounding the black 
hole \citep{DiMatteo:2005, Springel:2005f,Sijacki:2007, Vogelsberger:2013}. The heating effect of AGN feedback substantially reduces 
the gas density within the host halo \citep{vanDaalen:2011}. The gas expelled from the halo enriches the circum-galactic medium 
with metals \citep{Tescari:2011, Suresh:2015}. However, we shall show that the effect of enrichment due to AGN feedback on the 
observations we consider is relatively small. 
This is partly because the larger haloes hosting AGN are relatively rare at $z=2-4$, and 
partly because the ejected gas is heated beyond the temperature where it can produce a strong \civ~absorption signal.

Gas is enriched both by nearby star particles and by supernova winds.
Enrichment events are assumed to occur from asymptotic giant branch stars and supernovae, with the formation rate of each calculated 
using a \cite{Chabrier:2003} initial mass function. The elemental yields of each type of mass return are as
detailed in \cite{Vogelsberger:2013}. Metals are distributed into the gas cells 
surrounding a star using a top-hat kernel with a radius chosen to enclose a total mass equal to $256$ times 
the gas element mass targeted by the refinement algorithm \citep[see ][]{Vogelsberger:2013}. 
This models the expansion of the metals from the initial stellar event during the first timestep.
To ensure that our results are not sensitive to this parameter, 
we ran a simulation where metals were distributed within a radius enclosing $16$ times the 
gas element mass, with minimal changes to our results. Advection apparently erases the effect of the enrichment radius 
on short time-scales.

\subsection{Artificial Spectra}
\label{sec:spectra}

To compute the absorption profile along sightlines in our simulations, we need to calculate the ionization fraction of carbon ions 
within each gas cell. We use a lookup table generated by running \cloudy, version 13.02 \citep{Ferland:2013}, as described in \cite{Bird:2014a} and publicly 
available at \url{https://github.com/sbird/cloudy_tables}.

\cloudy~is run in single-zone mode, which assumes that the density and temperature are constant within each gas element.
We thus neglect density and thermal structure on scales smaller than the resolution limit of each simulation, which is consistent 
as this structure would be unresolved by definition. When generating simulated spectra, the ionization fraction is computed for each 
gas cell using the lookup table.

We assume ionization equilibrium and the presence of a uniform radiation background following \cite{Faucher:2009}.
We have checked explicitly, using a separate simulation where the amplitude of the radiation background was increased 
by a factor of two, that our results are insensitive to the background amplitude within current observational uncertainties.

\cite{Oppenheimer:2013} examined the effects of non-equilibrium cooling on the \civ~fraction and we have checked using the 
ionization tables they provide that the difference in \civ~fraction between equilibrium and non-equilibrium ionization 
for gas with $Z \sim 0.3 Z_\odot$ at $z=2$ is $\sim 10\%$, negligible for our purposes. This overestimates the impact; almost all 
the absorbing gas is at lower metallicity where the effect is much smaller. We neglect the impact of local ionizing radiation from stellar 
sources. \cite{Suresh:2016} showed that for a $10^{12} M_\odot$ halo 
this has an effect only $< 50$ kpc from the galaxy at $z=0.2$. Figures \ref{fig:civhalomass} and \ref{fig:civhalodist} shows that most of our absorbers are at distances $ > 100$ kpc from the center of a halo, 
and found near halos with $ < 10^{11} M_\odot$, so we do not expect local sources to have a significant effect.

We account for shielding from the radiation background at high hydrogen column densities, as described in \cite{Bird:2014a}. 
We have checked that this has no effect on \civ, but substantially increases the abundance of \cii. 
We model the onset of shielding using the fitting function from \cite{Rahmati:2013a}, who used a radiative transfer scheme to estimate
the threshold hydrogen density for gas to become neutral. This fitting function is implemented in \arepo, so that the effect of self-shielding 
on hydrogen cooling is included in the dynamics of the simulation. A rule of thumb is that gas becomes neutral at $\gtrsim 0.01$ cm$^{-3}$ with our redshift $2$ UVB.
Shielding for metal species is reduced at high energies by a frequency dependent correction to model the reduced cross-section of hydrogen 
photoionization to high photon energies, as explained in \cite{Bird:2014a}. 

The thermodynamics of dense, star-forming gas is set by the subgrid model for star formation \citep{Springel:2003}. This assumes that star-forming 
gas has a temperature $\sim 10^4$K and thus contains negligible \civ~and substantial \cii. Higher resolution simulations could reveal 
shocks in this dense gas, heating some of it to a temperature closer to the $10^5$K necessary to form \civ~via collisional ionization, and increasing the global abundance of 
high column density \civ~systems. Modelling this effect is beyond the scope of our present paper, but the relatively small covering fraction of star-forming gas 
suggests that changes in the \civ~absorption statistics are small.

The effective aperture of a physical quasar sightline, $< 1$ pc, is much smaller than the maximum spatial resolution of our simulations, 
$\sim 1$ kpc. We neglect any observational effects which may arise due to spatial structure on these small scales.
However, we shall show that the carbon absorption we consider here is produced in warm, diffuse circumgalactic gas which does not 
exhibit substantial structure on small spatial scales. While dense spatially compact clumps unresolved by our simulations may exist, 
we expect them to be confined to the high density inner regions of galaxies, where they can be sufficiently shielded from the UVB to collapse.
They would thus have a small covering fraction. We acknowledge, however, that recent observations \citep[e.g.][]{Crighton:2015}, may suggest 
otherwise, but resolving this problem is beyond the scope of this work.

Our artificial spectra are generated as described in detail in \cite{Bird:2014a}, using the implementation available at \url{https://github.com/sbird/fake_spectra}. 
Once the mass of the desired ion in each gas element has been computed, it is interpolated onto a sightline using a Voigt profile convolved with an SPH kernel.
The $b$-parameter of the Voigt profile is set by the temperature of the particle, naturally including thermal broadening. 
Each gas element is redshifted by its velocity parallel to the sightline, naturally accounting for peculiar velocities. Generated sightlines are noiseless. The 
size of a spectral pixel is $5$ km/s when comparing to \cite{DOdorico:2010} and \cite{Cooksey:2013} and $50$ km/s when comparing to \cite{Rubin:2014}.

We compute column densities primarily by interpolating the mass of the desired ion in each gas element to the sightline, as for the optical depth. 
Unlike when computing the optical depth, we do not redshift the particle according to its peculiar velocity, and use a top-hat spherical kernel rather than a Voigt profile.
Thus our column densities correspond to the integrated physical density field in the simulation, and allow us to examine the physical state of the absorber. 
This differs from common observational practice, which uses a Voigt profile fit from the optical depth. However, we have checked with a subset of our simulated spectra that performing
Voigt fitting on the artificial spectra to estimate the column densities does not affect our results.



\section{Absorber Properties}
\label{sec:absorbers}

\begin{figure*}
\includegraphics[width=0.45\textwidth]{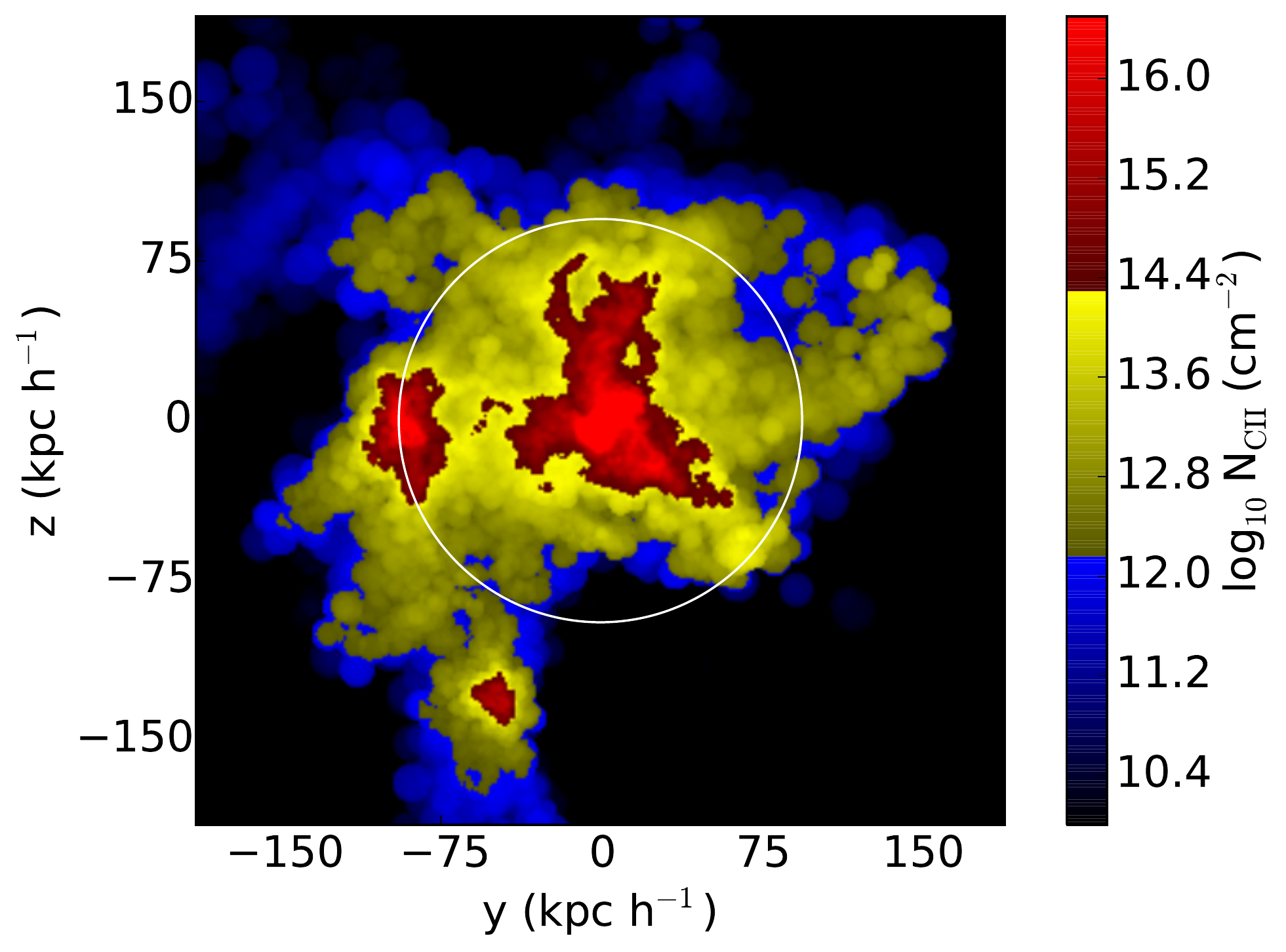}
\includegraphics[width=0.45\textwidth]{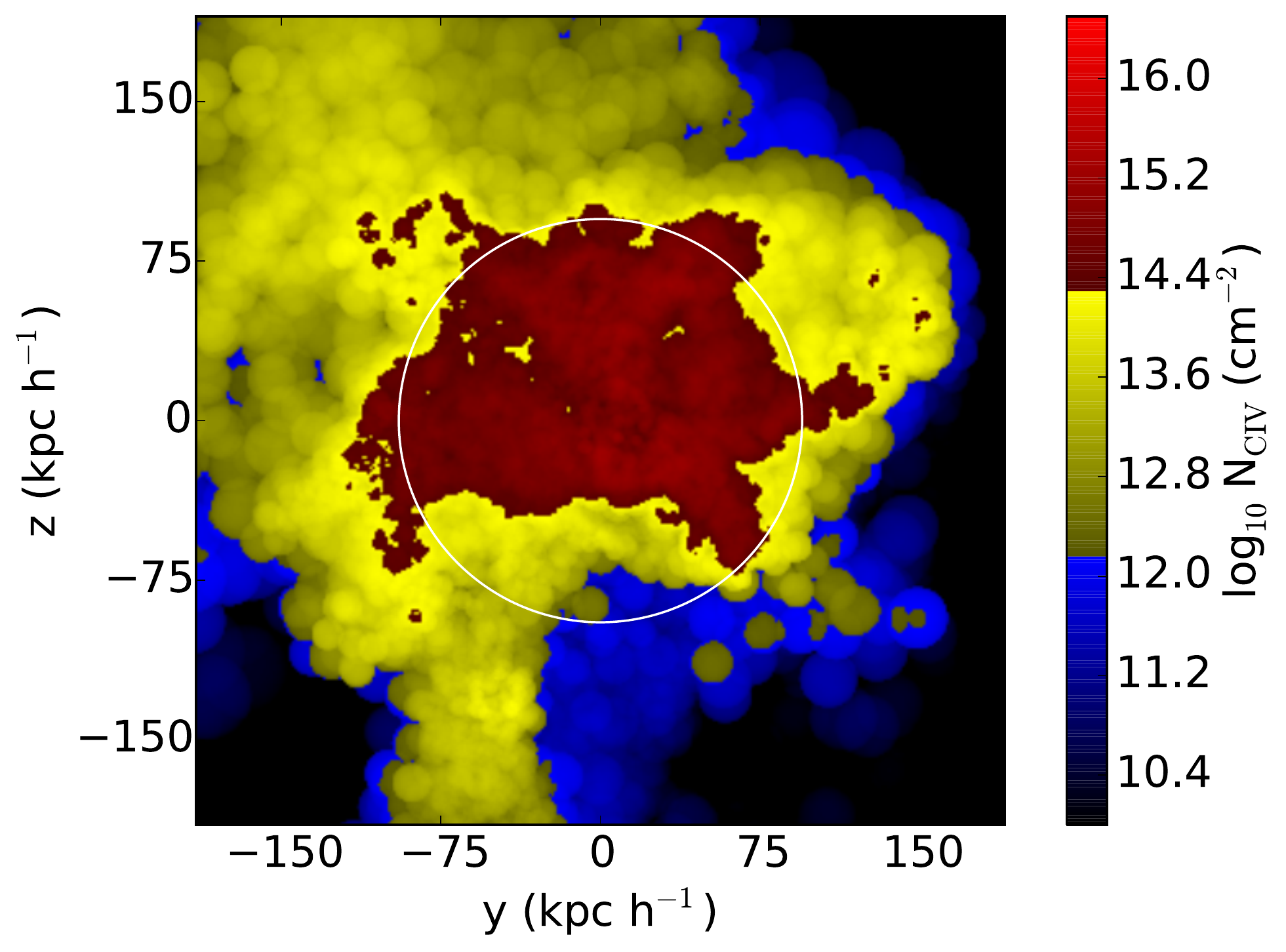}\\
\includegraphics[width=0.45\textwidth]{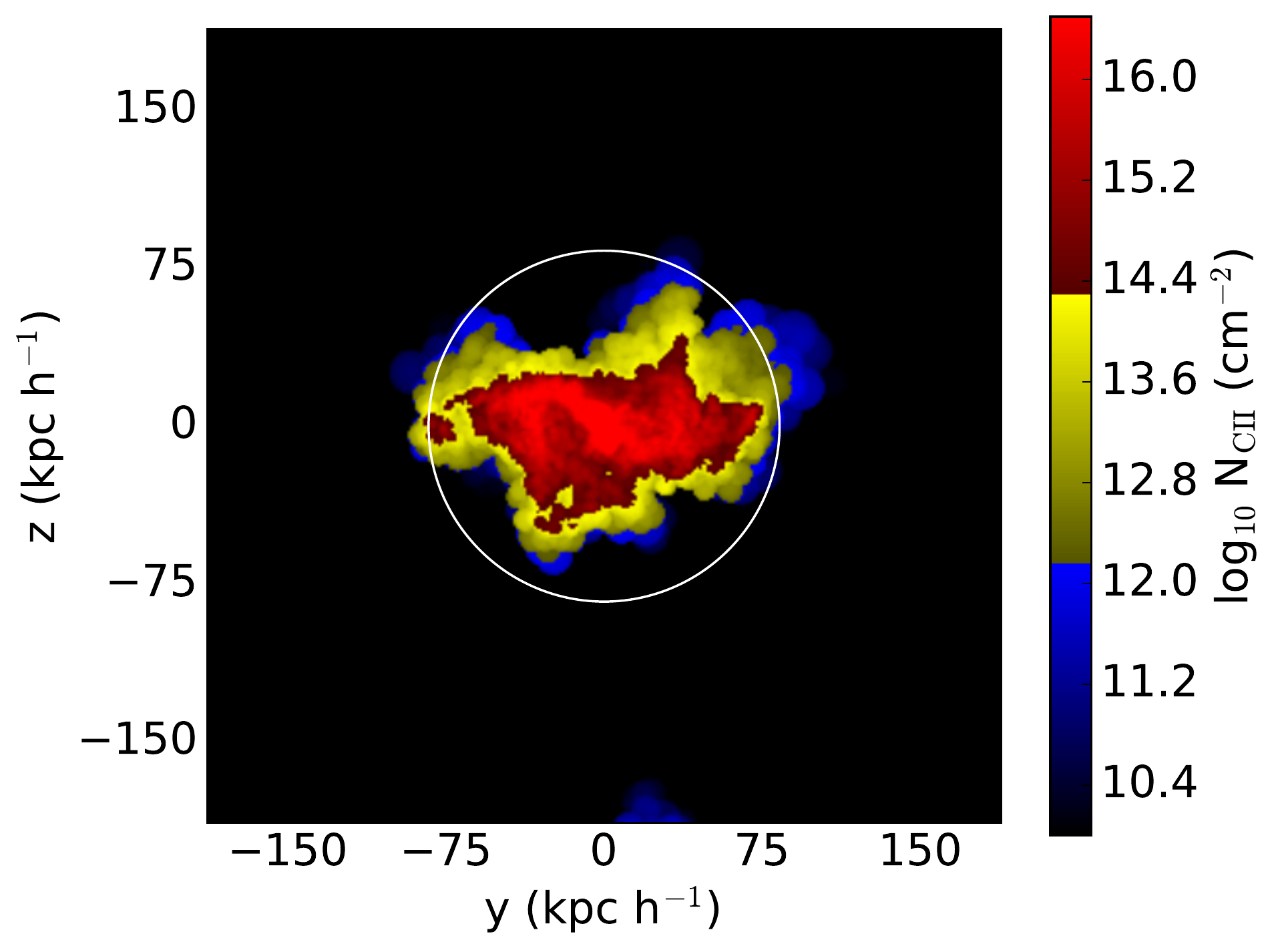}
\includegraphics[width=0.45\textwidth]{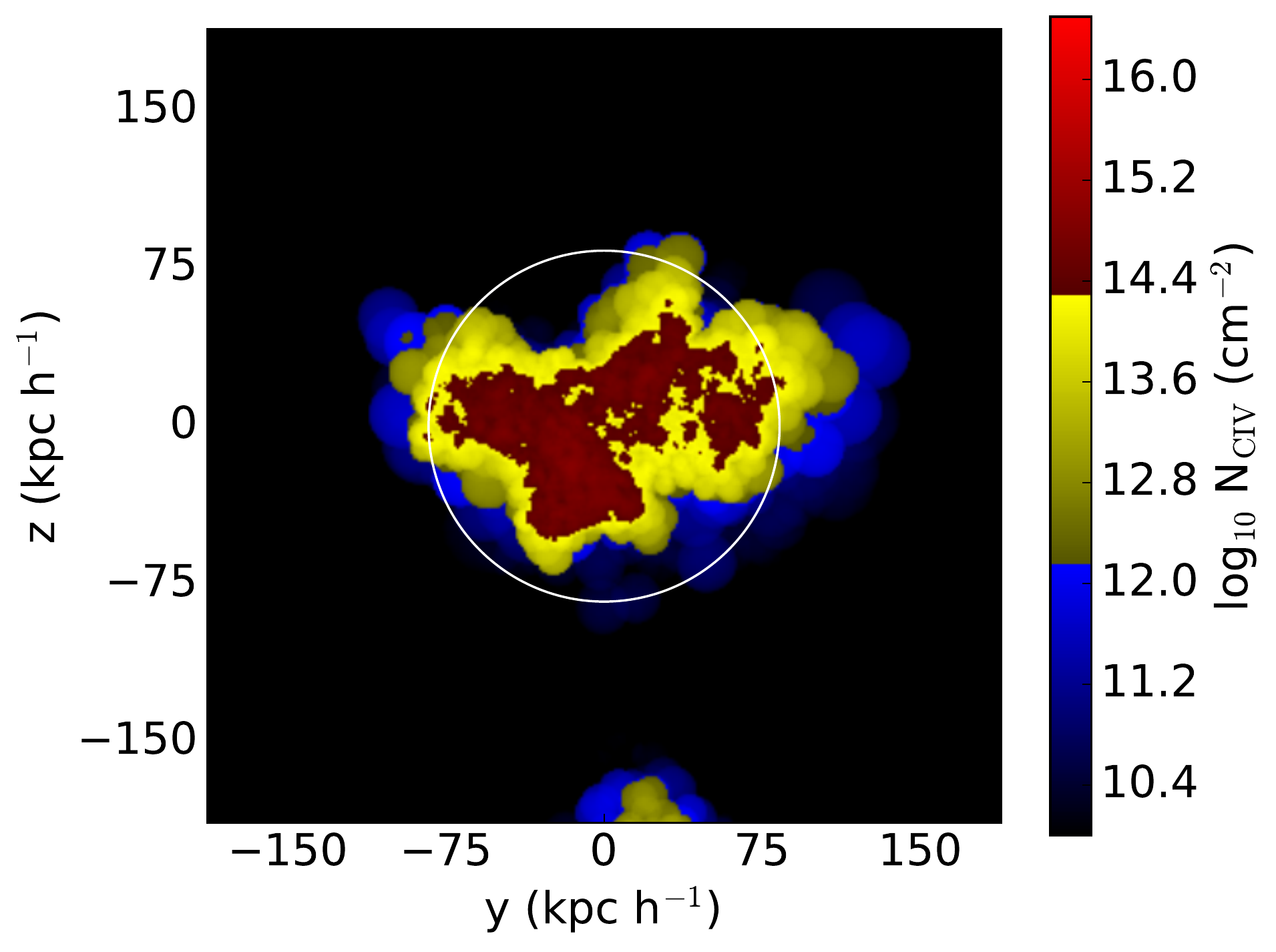}

\caption{Top: Column densities around a typical DLA-hosting halo in the WARM simulation at $z=2$.
Bottom: Column densities around a typical DLA-hosting halo in the ILLUS 25 simulation at $z=2$.
The halo has mass $\approx 6 \times 10^{10} \Msun$ and the white circle marks the halo virial radius.
Left: \cii~column density. Right: \civ~column density. In WARM, both \civ~and \cii~are present at detectable levels throughout the halo, up to the virial radius.
In the fiducial model, there is a lack of weak absorbers as the winds do not enrich gas outside the halo.
The circular features are artefacts of our spherical projection kernel.
}
\label{fig:civhaloessmall}
\end{figure*}

\begin{figure*}
\includegraphics[width=0.45\textwidth]{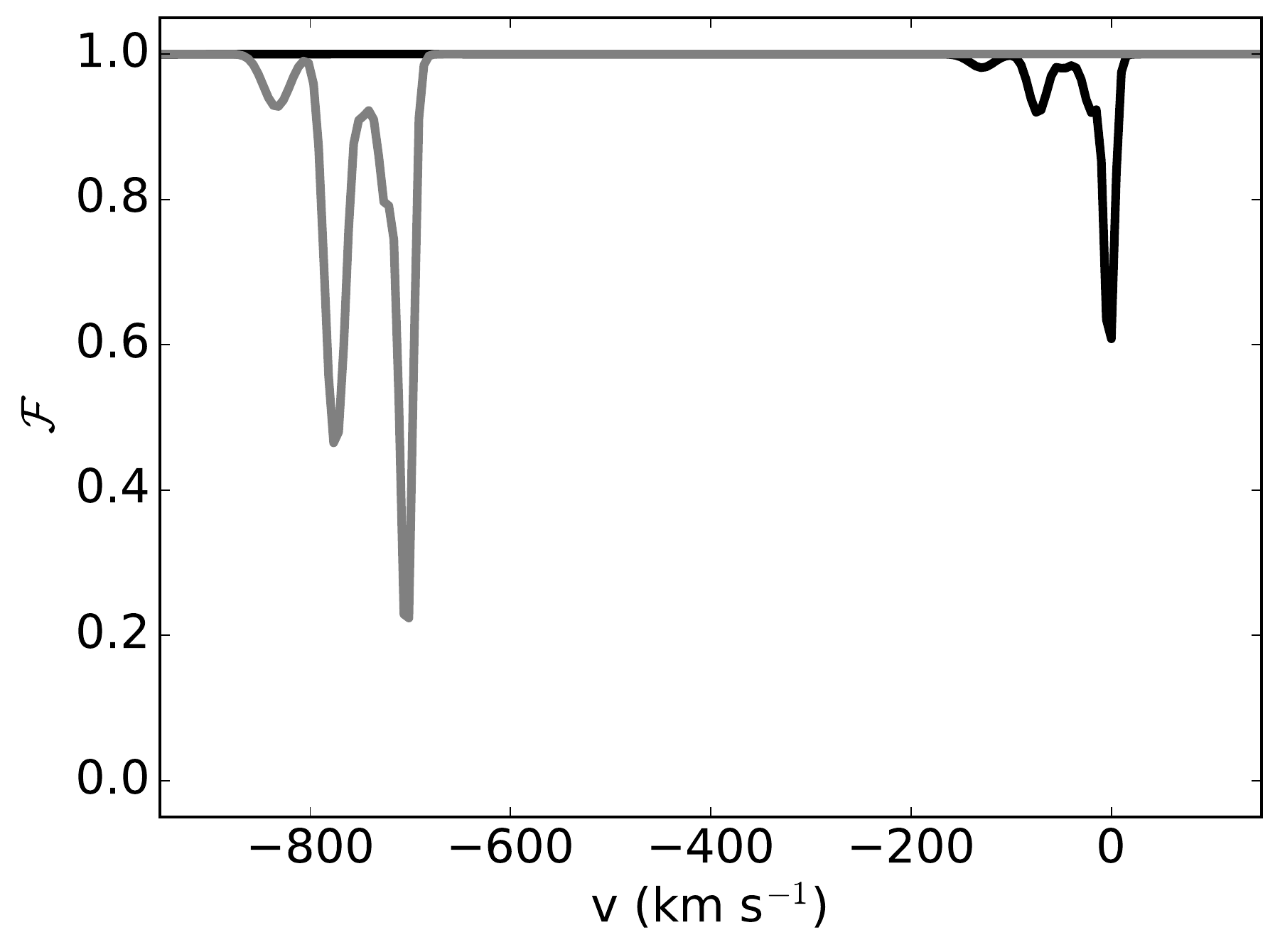}
\includegraphics[width=0.45\textwidth]{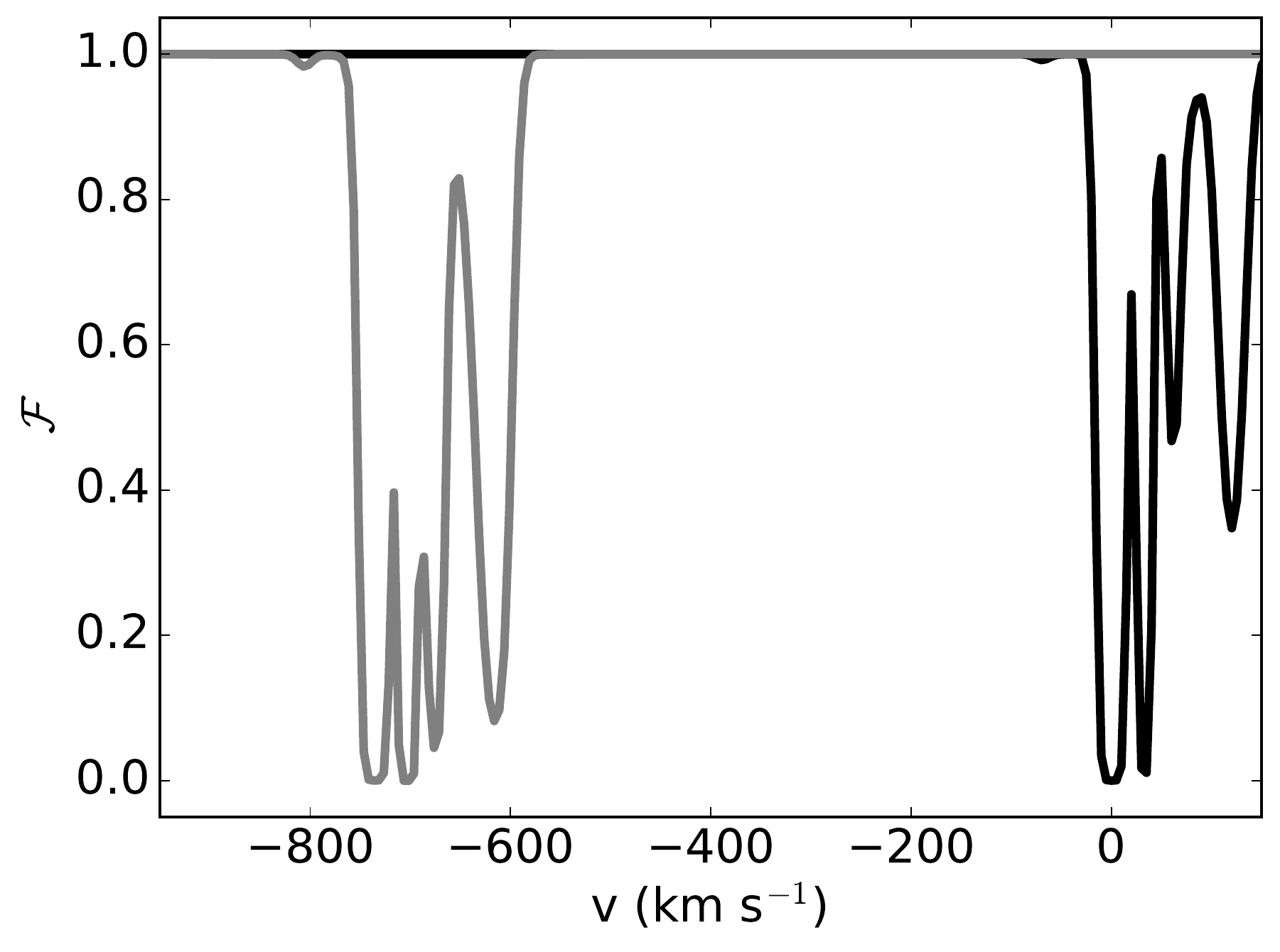}\\
\caption{Two example \civ~spectra, from the WARM simulation at $z=2$. The $1550$~\AA~line is shown in black, while the $1548$~\AA~doublet 
is overlaid in grey. (Left) an unsaturated absorber. (Right) a saturated line with a maximum optical depth $\tau > 4$.
}
\label{fig:civspectra}
\end{figure*}

Figure \ref{fig:civhaloessmall} shows the column density in \cii~and \civ~around a typical simulated halo in both our warm winds and Illustris simulations, with 
a mass of $6 \times 10^{10} \Msun$ at $z=2$. The column density is calculated by projecting the mass in each particle using a spherical top-hat kernel 
onto a gridded cube $300 \kpch$~across and centred on the halo. 
We have split the colour of \civ~absorption by column density thresholds. The transition between yellow and blue at $10^{12}$\NHunit~marks 
the lowest column density observable by \cite{DOdorico:2010}. The transition between red and yellow marks 
$W_{1548} \approx 0.2$ \AA~($N_\mathrm{CIV} \approx 2\times 10^{14}$ \NHunit), the 
detection threshold in \cite{Rubin:2014}. In this halo, the strongest \cii~absorbers are associated with dense clumps of cool material,
also producing strong HI absorption. Observable \cii~absorbers are present at similar levels in both simulations, but \civ~absorption is significantly more extended in the WARM simulation.

Figure \ref{fig:civspectra} shows two examples of simulated \civ~spectra from our WARM simulation, and we have examined several more \civ~absorption profiles by eye. 
We chose these examples to illustrate that our \civ~spectra sometimes contain considerable structure. While weaker absorbers tend to have simple Gaussian 
profiles, \civ~systems with column density $> 10^{14}$ \NHunit~are especially complex. Absorption profiles often contain multiple features, sometimes overlapping and 
in other cases separated by a few hundred km/s. This is mirrored in real spectra at high enough resolution \citep[e.g~][]{Haehnelt:1996}, and 
in some of the stronger systems in \cite{DOdorico:2010}. However, the relatively low resolution of the SDSS spectra from which 
the observed strong absorber sample is drawn produces smoothed profiles \citep{Cooksey:2013}.

\subsection{Host Halos}

\begin{figure}
\includegraphics[width=0.45\textwidth]{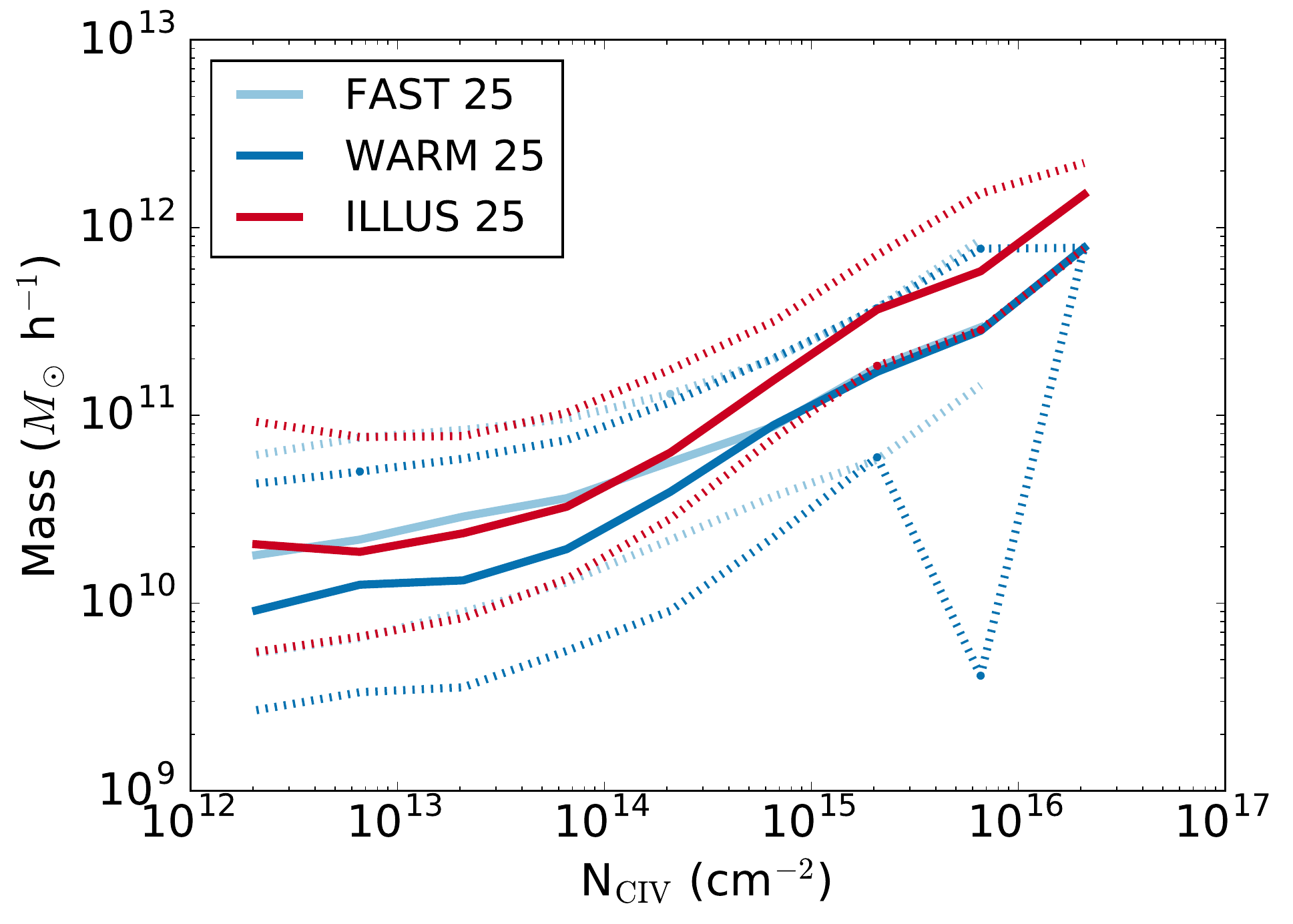}
\caption{Distribution of associated halo mass for different \civ~column densities at $z=2$ in the Illustris and WARM winds simulations.
Solid lines show the median in each column density bin, while dotted lines show the upper and lower quartiles. 
Strong absorbers, while extremely rare, are found predominantly in the largest halos, while weak absorbers are associated 
with a wide range of halos.
}
\label{fig:civhalomass}
\end{figure}

\begin{figure}
\includegraphics[width=0.45\textwidth]{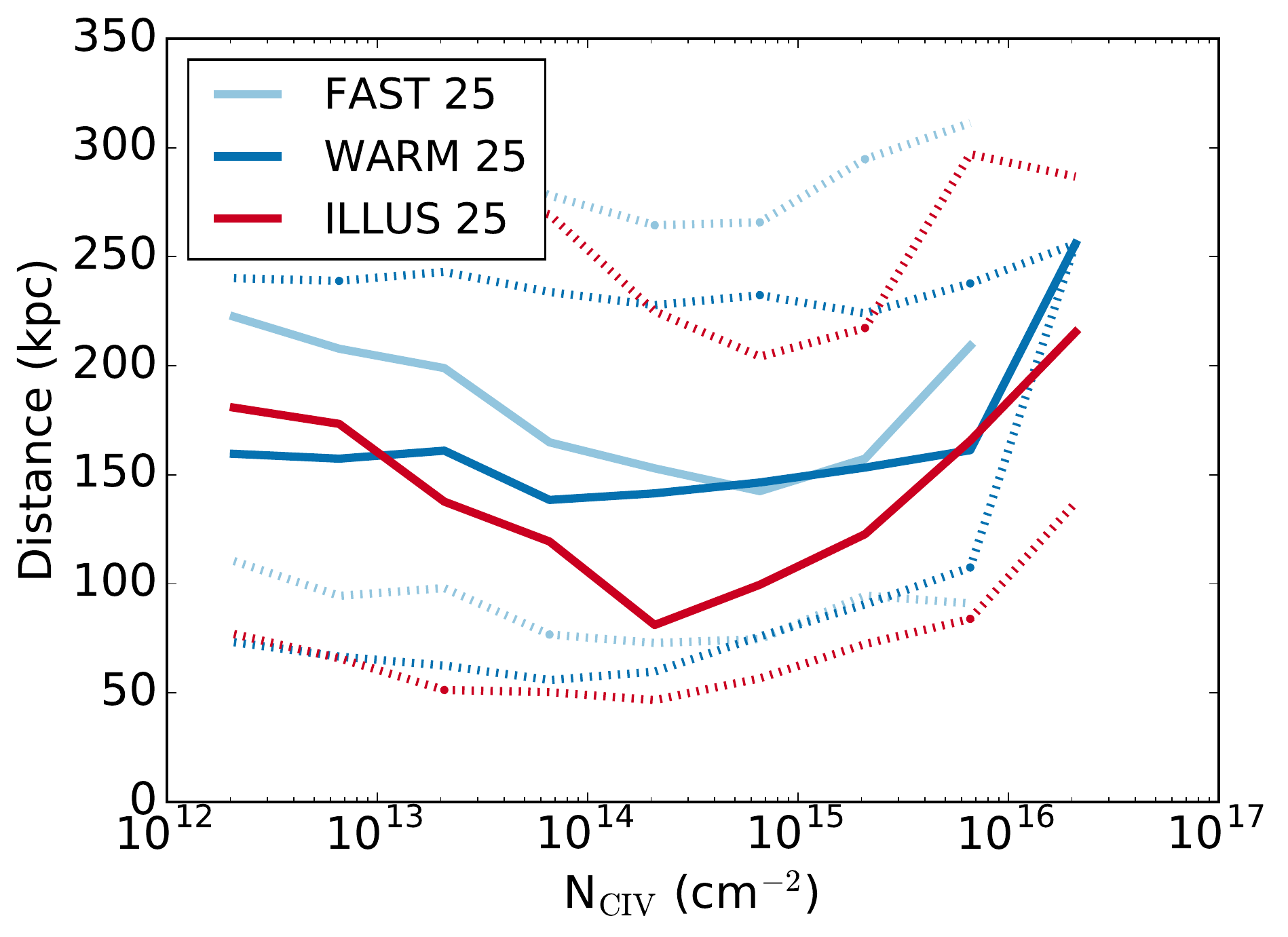}
\caption{Distribution of distances (in proper kpc) of \civ~absorbers from their assigned host haloes at $z=2$ in the Illustris and WARM winds simulations.
Solid lines show the median in each column density bin, while dotted lines show the upper and lower quartiles.
}
\label{fig:civhalodist}
\end{figure}

In \cite{Bird:2014} we associated DLAs to the dark matter halo in which they were found.
However, a substantial fraction of $N_\mathrm{CIV} > 10^{12}$ \NHunit~absorbers in our simulations at $z=2$ are located outside of any halo.
Since gas is enriched by stellar activity, one solution would be to assign an absorber to the halo which contains the 
majority of the stars which caused the enrichment. However, in our moving mesh simulations material is allowed to move 
between neighbouring gas cells, and so it is difficult to track the ultimate source of metals. 
Illustris contains an implementation of stochastic tracers designed to solve this problem \citep{Genel:2013}.
Unfortunately, we found that because these tracers follow the total gas mass, they did not provide adequate sampling of the metal flow.
Re-running Illustris with metallicity tracers is beyond the scope of our current paper.

Instead we assign absorbers to the largest halo within $400$ physical kpc, chosen as the minimum distance which assigned 
a halo to $99\%$ of absorbers. This matches the practice of observers, who often connect absorbers with the brightest object within some physical distance.
We make the approximation that the brightest object corresponds to the most massive halo, which is reasonable at 
these redshifts where relatively few galaxies are quenched. An alternative would be to weight 
distances by the virial radius of the halo concerned. This gives similar results, but given the observational difficulty 
of determining virial radii we decided against it. Reducing the maximum distance naturally reduces the mean distance from the absorber somewhat 
but does not change trends, as long as a threshold $\gtrsim 250$ kpc is chosen. Interestingly, this is roughly the mean distance between 
$10^{16}$ \NHunit~absorbers and their host halo.
The redshift of an absorber along a sightline is given by the mean \civ~mass-weighted redshift.
Our prescription will generally prefer larger haloes, which dominate the dynamics of a group, and, at $z=2-3.5$, the global star formation rate.
Many \civ~absorbers are also close to small haloes; with a low star formation rate, these haloes should not 
substantially affect overall enrichment, so our halo assignment procedure ignores them by design.

Figure \ref{fig:civhalomass} shows the mass distribution of \civ~absorber host haloes at $z=2$. 
Absorber column densities are computed as described in Section \ref{sec:civglweak}. Most weak \civ~absorbers are found around small haloes, which
are more common, but weak absorbers are also found around larger halos.
By contrast, absorbers with $N_\mathrm{CIV} \gtrsim 10^{15}$ \NHunit~are found only in haloes with mass $> 10^{11} \Msun$. 
Figure~\ref{fig:civhalodist} shows the mean distances of absorbers from the selected host halo, 
emphasising that absorbers of all strengths can be found at distances much larger than the virial radius of the halo 
we associate them with, which ranges between $300$ and $50$ kpc. 
The large distances found in these figures show that the more energetic winds are especially able to launch material further from the host halo, and thus 
produce \civ~absorption at larger distances. Indeed, the energy per unit mass of the wind ejecta in many cases exceeds 
the potential energy of the halo. This spreads the material over a wider region and we shall see that it substantially 
boosts the abundance of \civ~absorbers. The increased enrichment also means that strong absorbers can occur around 
smaller haloes than in the Illustris simulation.
Note there is large scatter in the halo-absorber distance, as \civ~absorption is produced by gas at a range of densities 
in a variety of spatial locations, rather than being confined to dense self-shielded gas at the center of a halo.


%
\begin{figure*}
\includegraphics[width=0.45\textwidth]{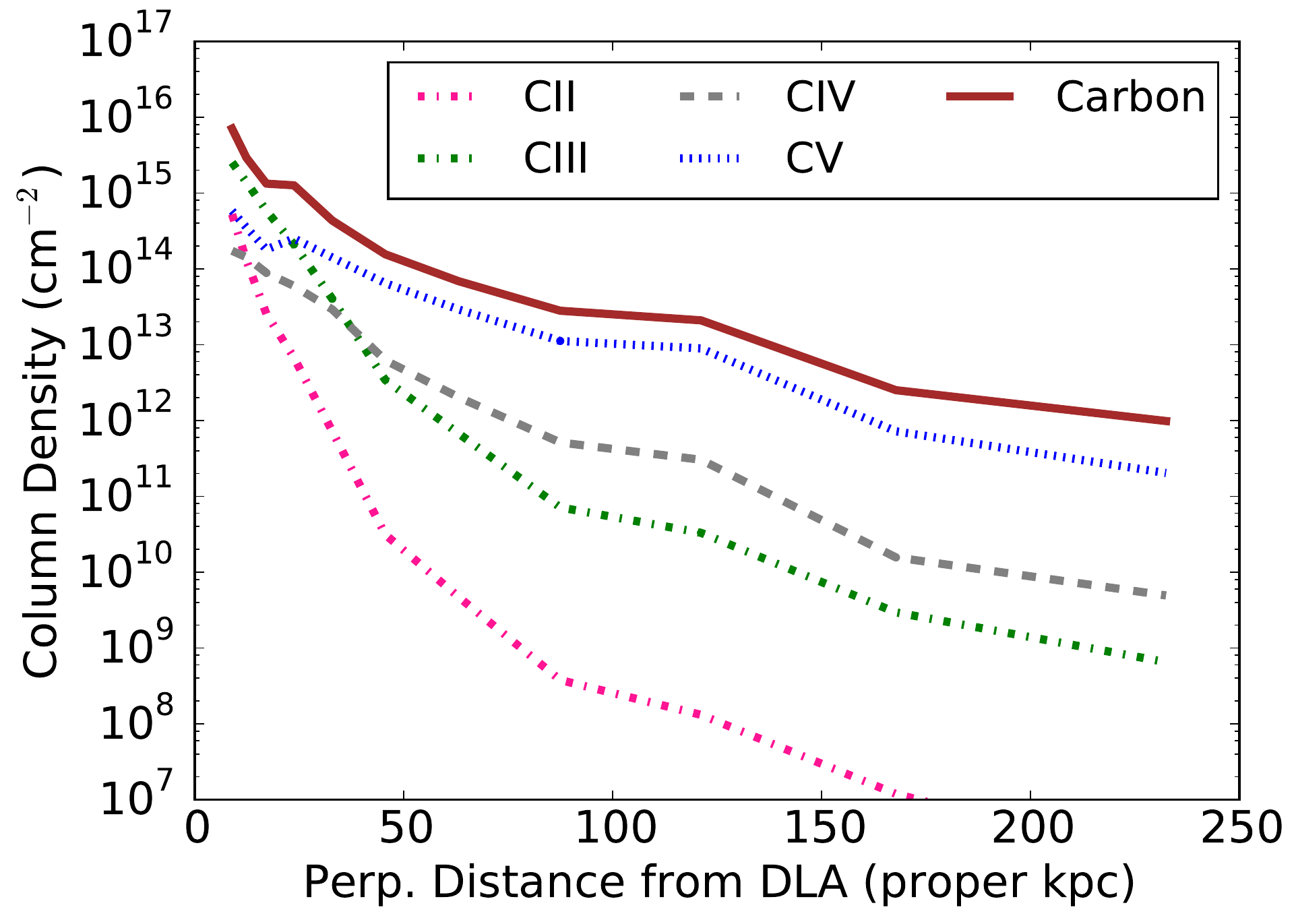}
\includegraphics[width=0.45\textwidth]{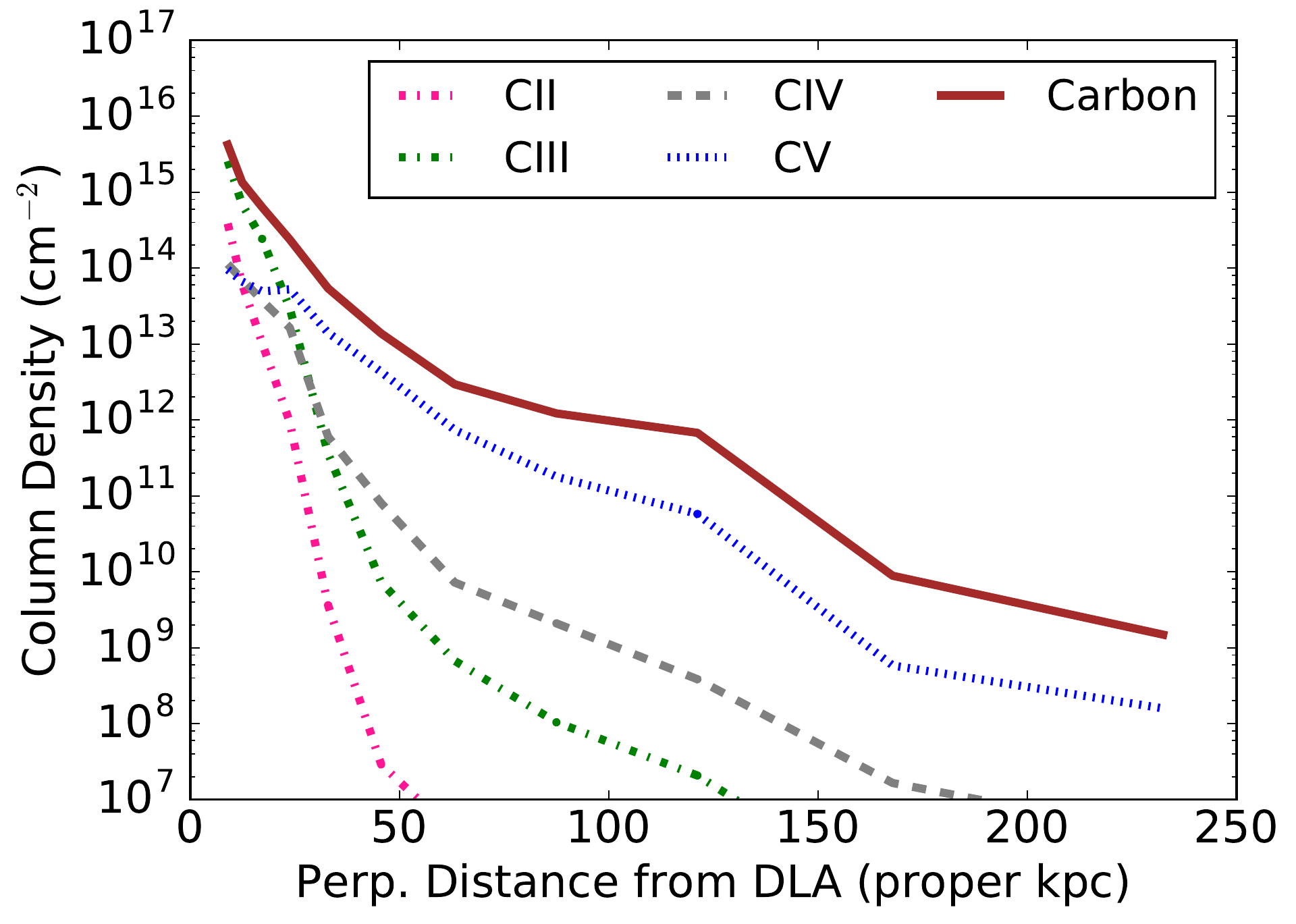}
\caption{Median column density for carbon ions as a function of perpendicular distance between paired sight-lines, one containing a DLA. 
Sight-line pairs are selected following \protect\cite{Rubin:2014}, as explained in the text. Left: the WARM winds $25 \Mpch$ simulation. 
Right: the Illustris $25 \Mpch$ simulation. Near the DLA the gas 
is dominated by \cii~and CIII, while further away CV is the most common ion. As the column density shows the density integrated along the 
sightline, some ions are at slightly different redshifts than the DLA. Because \civ~and CV are more often separated from the DLA, projection effects 
may boost their calculated abundances near the DLA relative to \cii. We compare to observations in Section \protect\ref{sec:civdlas}.
}
\label{fig:ionfrac_colden}
\end{figure*}

To aid interpretation of our comparison to the results of \cite{Rubin:2014} in Section \ref{sec:civdlas}, 
Figure \ref{fig:ionfrac_colden} shows the median radial carbon column density profile for our simulated DLAs.
The column density is shown as a function of the perpendicular distance between paired sight-lines. 
Column densities are computed by integrating the density along the non-DLA sightline within $300$ km/s of the DLAs peculiar velocity.
Sight-line pairs are chosen so that at least one of each pair contains a DLA, but no other condition is placed on the material between the sight-lines. 
The shown profiles are averaged over $z=2-3$. These choices model the selection function used in \cite{Rubin:2014}, and are discussed further in Section \ref{sec:civdlas}.
Less than $20$ kpc from the DLA, the most common ionization state is CIII. Near the DLA the gas is cool and dense and absorbs in HI.
Further away from the DLA the dominant ionization state changes to \civ~and CV. CV is not easily observable in absorption, 
so this phase is observed most easily as \civ.

The density profile is dependent on the wind model. FAST is omitted as its profile strongly resembles WARM. 
Profiles in WARM and FAST are flatter than in Illustris. The effect of the increased energy in the WARM winds simulation is to boost the 
abundance of all carbon ions more than $30 \kpch$ from the DLA, although only the increase in \civ~is observable. As discussed further in Section \ref{sec:civdlas}, this 
is due to the higher energy winds, which are able to launch metals out of the star-forming regions where they originate more easily.

\subsection{Ionisation balance}
\label{sec:ccioniz}

\begin{figure}
\includegraphics[width=0.45\textwidth]{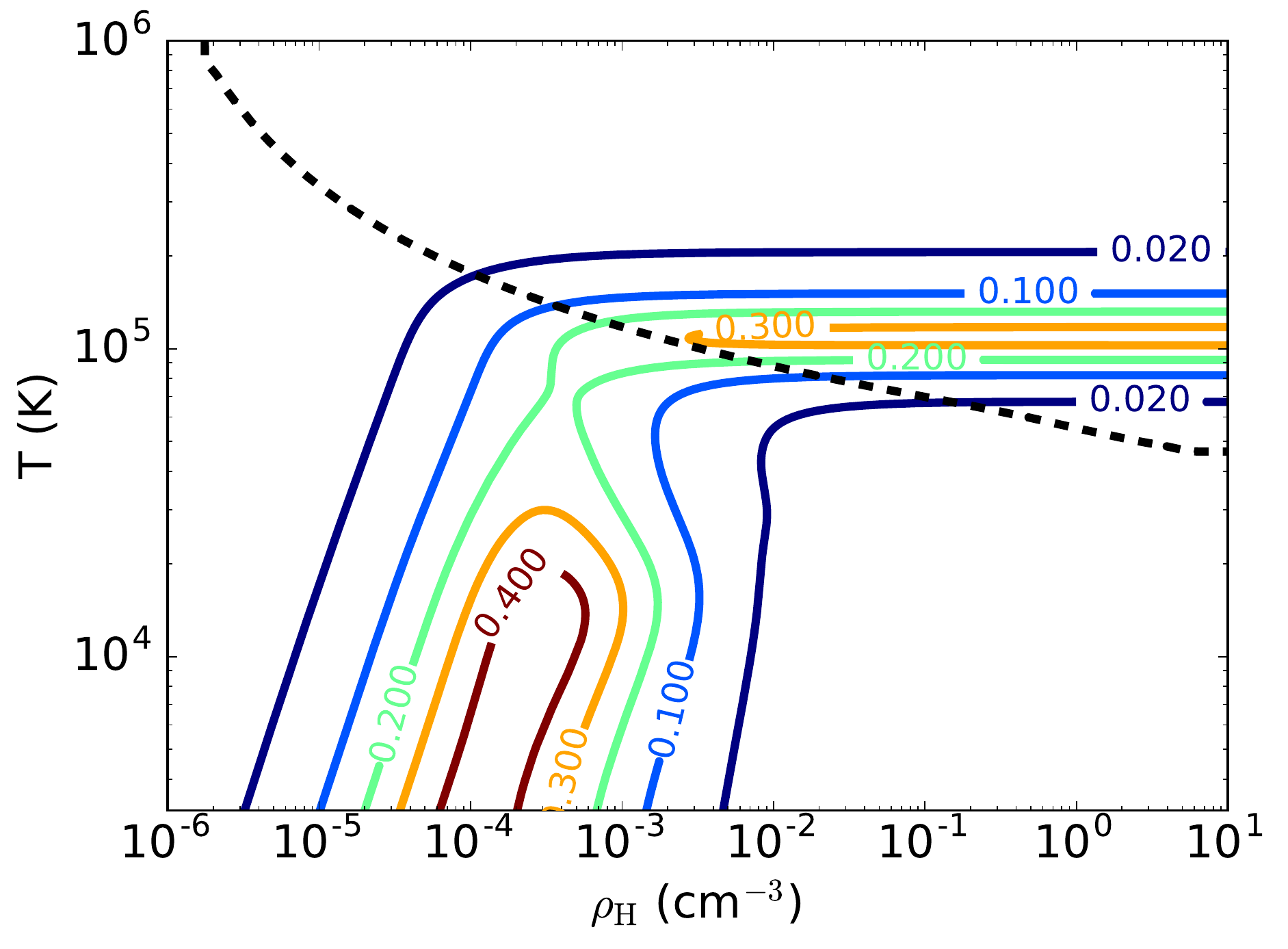}
\caption{Contours show regions of the temperature-density plane at $z=3$ where the mass fraction of 
carbon in \civ~is above $0.01$, $0.1$, $0.2$, $0.3$ and $0.4$, respectively. \civ~occurs in a collisionally 
ionized phase at $T \sim 10^5$K and a photo-ionized phase at $\rho \sim 10^{-4}$ cm$^{-3}$.
The black line shows the boundary between the two phases: collisional ionization dominates above the line, photo-ionization below it.
At $z=2$ the maximum \civ~fraction in the photo-ionized region is increased by approximately $20\%$, due to the increased UVB amplitude.
}
\label{fig:ionfrac}
\end{figure}

Figure \ref{fig:ionfrac} shows the \civ~ionization fraction for a range of temperatures and densities. At temperatures $T \sim 10^5 $ K and 
densities $\rho > 10^{-3}$ cm$^{-3}$, \civ~is produced by collisional ionization, while at $T < 20000$ K and $\rho = 10^{-4} - 10^{-3}$ cm$^{-3}$
\civ~is produced by photoionization. The black dashed line in Figure~\ref{fig:ionfrac} shows the boundaries between these two regimes 
as a function of temperature and density. To compute this boundary, collisional and photo-ionization rates at each temperature 
and density are tabulated using \cloudy. The photo-ionization rate depends on the density and the amplitude of the radiation background, while the collisional 
ionization rate depends on the temperature and density of the material. Higher densities are collisionally dominated at lower temperatures, 
as the increase in particle density produces more collisions.

As we have access to the physical states of absorbers in the simulations, we are able to determine how much of the mass in \civ~absorbers 
is in each phase. We first compute whether each gas cell is in a collisionally or photoionized regime, based on its position
in the temperature-density plane. We then compute the \civ~column density of all collisionally ionized cells by integrating the \civ~density 
of each particle along the sightline, and divide by the total column density. We thus obtain the fraction of \civ~absorption from collisionally 
ionized sources in each absorber. One advantage of this definition is that it avoids projection and smoothing effects by using the 
temperature and density of the gas cell directly, rather than these quantities averaged over a spectral pixel. This is important because the 
ionization fraction depends non-linearly on the temperature. Thus the collisional ionization rate for the averaged temperature is not necessarily
close to the average of the collisional ionization rates from each particle.

\begin{figure}
\includegraphics[width=0.45\textwidth]{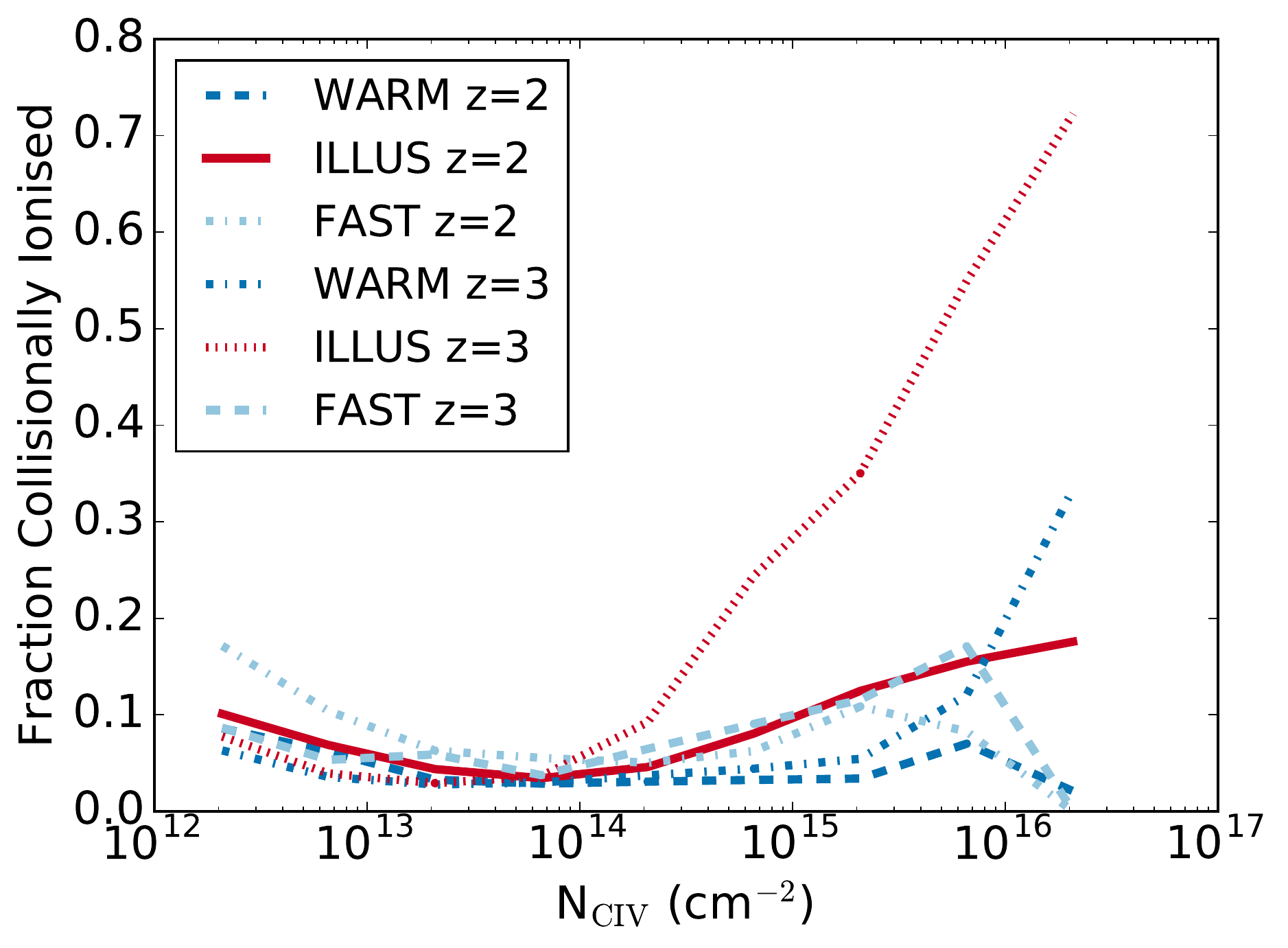}
\caption{The fraction of \civ~which is collisionally ionized as a function of column density. 
Shown are our three feedback models at $z=3$ and $z=2$. Note that although collisionally ionized material in some cases dominates 
the strongest absorbers, the abundance of these absorbers is low. Overall most \civ~absorbers are photo-ionized.
}
\label{fig:collisional}
\end{figure}

Figure \ref{fig:collisional} shows the fraction of \civ~which is collisionally ionized as a function of column density.
Overall, the collisional ionization fraction of \civ~is low: $\sim 4\%$ for WARM and $\sim 6\%$ for ILLUS.
This is because the number density of \civ~absorbers is dominated by the lower column densities, which are heavily photoionized.
Stronger absorbers have a higher collisional fraction, especially with the default Illustris wind model, and at $z=3$. This is 
because these strong absorbers are often associated with self-shielded gas, for which the photoionization rate drops to zero.
At higher redshift, the UVB amplitude is lower, and gas self-shields at a lower density, so a higher fraction of 
the absorbers are collisionally ionized.

At all redshifts, the warm winds simulation shows a substantially lower fraction of collisionally dominated absorbers. 
Most of the collisionally ionized absorbers are still present, but the warm winds model heats the gas, better enabling it to free-stream out of the halo, 
and leading to additional \civ~absorption. As this arises predominantly from low density photo-ionized material, it decreases the fraction of absorbers that are 
collisionally ionized. We shall see that this property of the wind model also allows it to better match observations, suggesting that, physically, 
even strong absorbers are largely photoionized.

OD06 found a moderately higher fraction of collisionally ionized \civ~absorbers than in our simulations. 
This appears to be due to our differing definitions of collisional ionization, and does not reflect a change in the physical
properties of the gas. OD06 use a simple temperature threshold of $10^5$K for densities above $10^{-5}$ cm$^{-3}$ and reduced temperature thresholds 
for lower densities, resulting in a much larger region of collisional ionization. We regard our definition as more physical. 
In particular, lower densities should require higher temperatures to be collisionally dominated.

\section{Comparison to Observations}
\label{sec:observations}


In this section we compare the results of our simulations to observations. We have created simulated \civ~measurements comparable 
to three observational surveys. Section \ref{sec:civglweak} compares to \cite{DOdorico:2010}, who observed weak \civ~absorption 
in randomly selected quasar sightlines. Section \ref{sec:civglstrong} compares our results to observations of strong absorber systems \citep{Cooksey:2013}.
Finally, section \ref{sec:civdlas} compares the \civ~and~\cii~absorption around DLAs (used as a tracer of the galaxy population) 
to the results of \cite{Rubin:2014}.

\subsection[CDDF of weak CIV]{Column Density Distribution of Weak \civ~Absorbers}
\label{sec:civglweak}

\begin{figure*}
\includegraphics[width=0.45\textwidth]{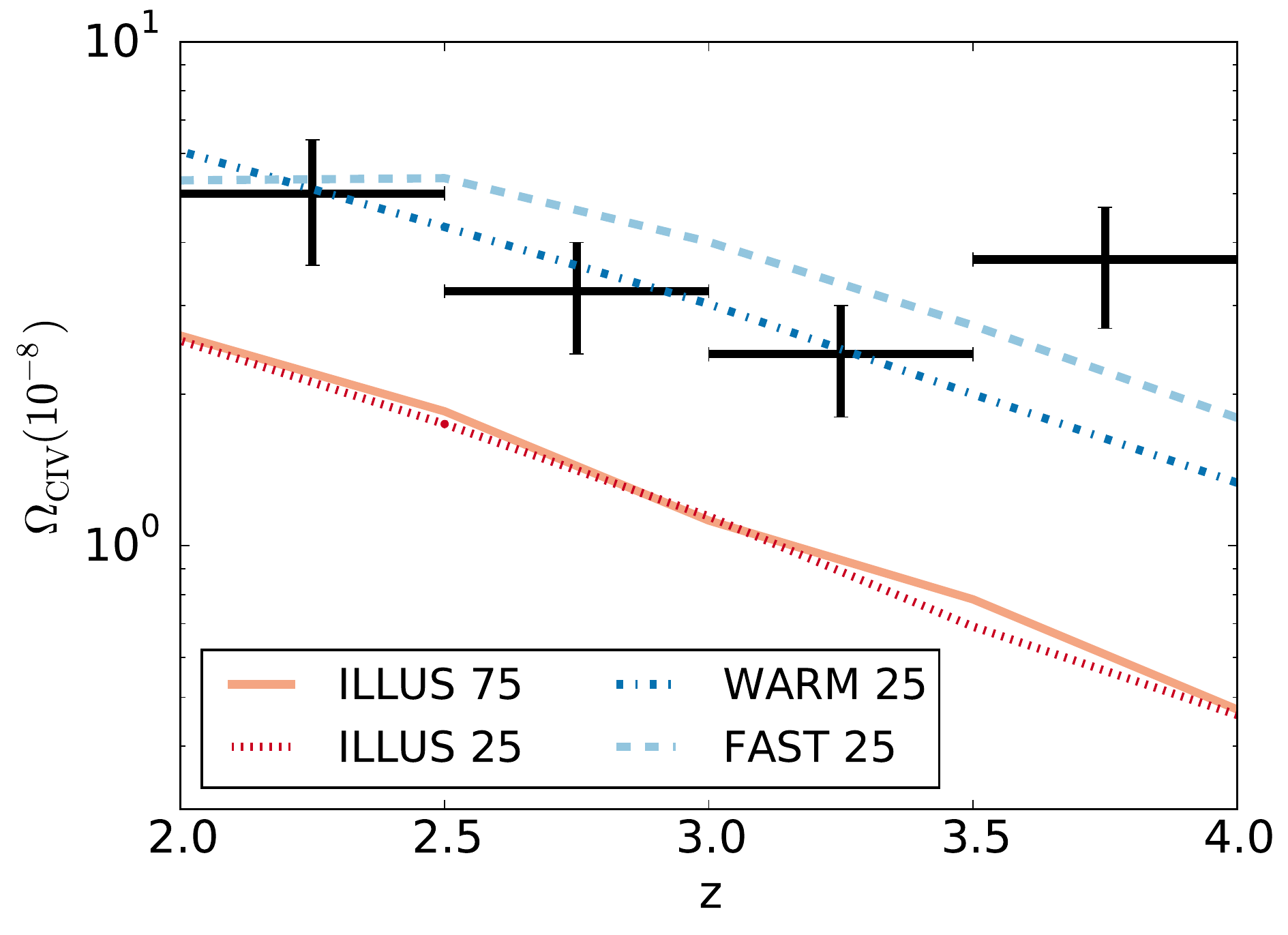}
\includegraphics[width=0.45\textwidth]{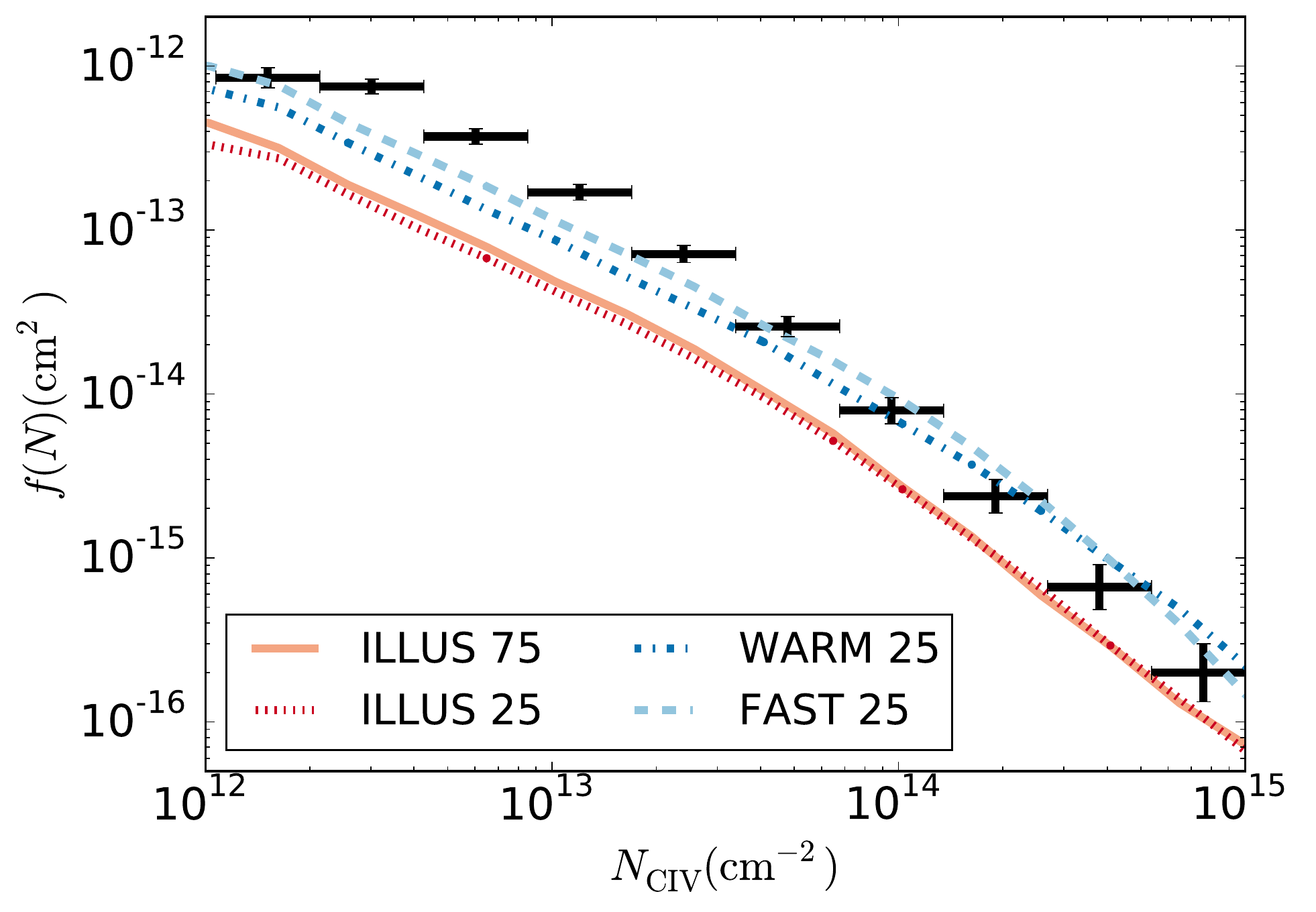}
\caption{Left: The evolution of $\Omega_\mathrm{CIV}$, the integrated \civ~CDDF between column densities 
of $10^{12}$ and $10^{15}$ \NHunit~in the simulations considered here. Right: The column density function, $f(N)$, of \civ~in 
the same column density range, averaged over redshifts $2.0 < z < 3.5$. Points with error bars show the measurements of \protect\cite{DOdorico:2010}.
The WARM and FAST models match observations substantially better than the Illustris model.
}
\label{fig:omega_civ}
\end{figure*}

We define the column density distribution, $f(N)$, such
that $f(N)$ is the number of absorbers per unit column density per unit absorption distance with column density 
in the interval $[N, N + {\rm d}N]$. Thus,
\begin{align}
 f(N) &= \frac{F(N)}{\Delta N \Delta X(z)}\,,
\end{align}
where $F(N)$ is the number of absorbers along simulated sightlines in a given column density bin. $\Delta X(z)$ is the absorption distance per sightline, 
defined to account for evolution in line number density with the Hubble flow:
\begin{equation}
 X(z) = \int_0^z (1+z')^2  \frac{H_0}{H(z')} {\rm d}z'\,.
 \label{eq:absdist}
\end{equation}
For a box of co-moving length $\Delta L$ we use the approximation $\Delta X = (H_0 /c) (1+z)^2 \Delta L$ \citep{Bahcall:1969}.

$\Omega_\mathrm{CIV}$ is the integral of the first moment of the CDDF:
\begin{equation}
 \Omega_\mathrm{CIV} = \frac{12 m_\mathrm{P} H_0}{c \rho_c}\int_{10^{12}}^{10^{15}} N f(N, X) \mathrm{d} N\,.
\end{equation}
$\rho_c$ is the critical density at $z=0$ and $m_\mathrm{P}$ is the proton mass. To ensure a fair comparison, the column density limits 
of $10^{12} - 10^{15}$ \NHunit~have been chosen to match those used in \cite{DOdorico:2010}\footnote{\cite{Rahmati:2015} integrated from $0$ to $\infty$. Illustris has more strong 
systems with $N_\mathrm{CIV} > 10^{15}$ \NHunit~than EAGLE, so for our simulations this would over-estimate the observed $\Omega_\mathrm{CIV}$ by a factor of $2-4$.}. 
The lower limit corresponds to the level below which their survey can no longer correct for completeness, while the upper limit is set by the small sample of 
higher column densities.

We generated a simulated catalogue using $10,000$ synthetic sightlines positioned at random through each simulation snapshot.
Sightlines with a \civ~column density less than $10^{12}$~\NHunit~were discarded until we had a sample of $10000$ sightlines containing 
observable \civ~absorbers. Average abundances included the path length for the discarded spectra, effectively treating them as non-detections. 
We compute column densities by summing densities over $50$ \kms~intervals along the synthetic spectra.
This differs from \cite{DOdorico:2010}, who measured column densities by fitting Voigt profiles to spectral lines and combining absorbers closer than $50$\kms. 
We have however checked that the difference between our direct summation of the density field and a full Voigt fit is small. We also checked that our 
results were unchanged when combining absorbers $150$\kms~apart. The absorption in most sightlines arises from gas in a relatively small spatial area,
although this gas frequently has a complex velocity structure which leads to a more dispersed absorption profile.

Figure \ref{fig:omega_civ} compares our simulations to the measurements of \cite{DOdorico:2010}. These 
include the column density distribution function (CDDF) of weak \civ~absorbers and the evolution of the matter density in \civ, $\Omega_\mathrm{CIV}$.
The CDDF measured by \cite{DOdorico:2010} is computed from absorbers over a wide range in redshift, $1.6 < z < 3.6$. 
To approximate the redshift distribution of the survey, we average the CDDF from different simulation snapshots at $z=2-3.5$, weighted by 
the fraction of the total path length in each redshift bin in the observational survey.
Note that, as other observational surveys have a different redshift distribution and selection function, it is not strictly correct to compare the
simulated \civ~CDDF shown in Figure \ref{fig:omega_civ} to the results of other surveys.

As shown in Figure \ref{fig:omega_civ}, all our simulations agree well with the shape of the CDDF, 
and evolve similarly with redshift. By comparing $75$ and $25 \Mpch$ boxes we show that, as expected from the results of OD06, a $25 \Mpch$ box 
is sufficiently large to model \civ~systems at these column densities.

However, the Illustris feedback model produces too little \civ~by a factor of about $2$, as measured both by $\Omega_\mathrm{CIV}$ and the amplitude of the CDDF. 
FAST and WARM, the more energetic wind models, both produce increased \civ~enrichment and thus agree better with observations of $\Omega_\mathrm{CIV}$ and match the CDDF
for \civ~column densities $> 10^{13}$~\NHunit. They still produce too few of the lowest column density systems. However, as shown in 
Appendix \ref{sec:resolution}, increasing the resolution of the simulation also moderately increases the CDDF for $N_\mathrm{CIV} < 10^{13}$~\NHunit, due 
to star formation in newly resolved halos. A higher resolution version of the WARM or FAST simulation may thus be in agreement with the observed faint end CDDF.

Both WARM and FAST drive more energetic outflows that allow the metals to more easily escape the host halo, and 
have similar wind energy per unit mass, shown by \cite{Suresh:2015} to be the feedback parameter with most influence on the metallicity 
of the circumgalactic medium. That they produce similar results suggests that the statistical properties of \civ~thus far observed are sensitive only 
to how much energy is transferred to the circumgalactic gas, not how it is transmitted. We have checked that the effect of AGN feedback 
is mild for these lower column density systems. 

OD06 and OD08 adopt winds with a mass loading $\eta \sim v_\mathrm{w}^{-1}$, where $v_\mathrm{w}$ is the wind velocity. 
Illustris on the other hand uses $\eta \sim v_\mathrm{w}^{-2}$, and so has a wind mass loading which increases more steeply in lower mass haloes. 
OD06 increase the mass loading in small haloes by allowing the wind speed to depend on metallicity, so that a lower metallicity induces a higher wind velocity. 
Overall this leads to an increase in energy per unit mass which boosts $\Omega_\mathrm{CIV}$ by about $50\%$ compared to Illustris. 
The difference between Illustris and OD06 is slightly smaller than that between Illustris and the observations, as \cite{DOdorico:2010} measures a CDDF with a 
larger amplitude than the earlier measurements \cite[e.g.][]{Songaila:2001} matched in OD06.

\begin{figure}
\includegraphics[width=0.45\textwidth]{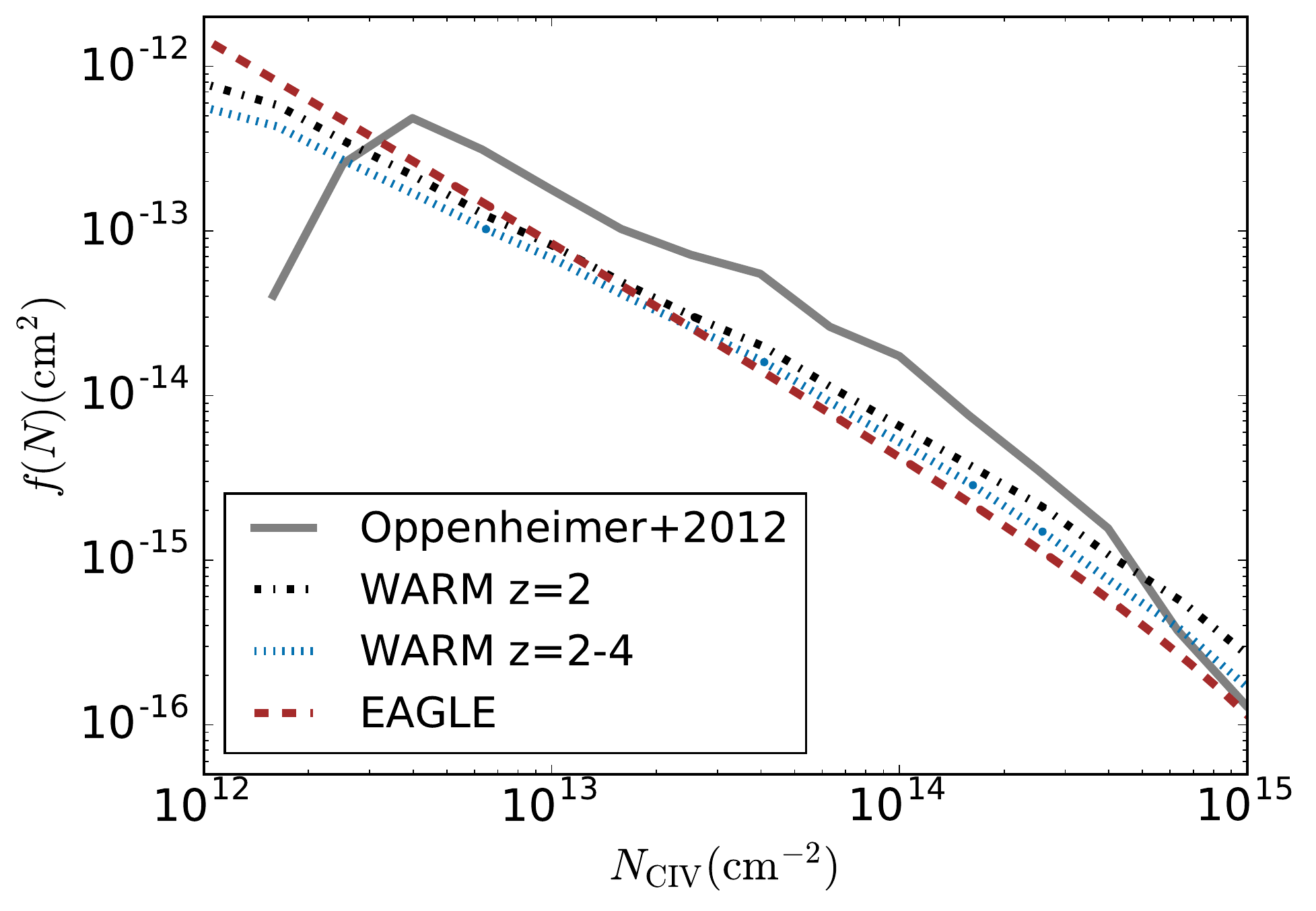}
\caption{The CDDF from our WARM model at $z=2$ (dot-dashed) compared to the EAGLE simulation at $z=2$ \protect\citep{Rahmati:2015} (dashed) and the redshift 
averaged $z=1.5-2$ CDDF from \protect\cite{Oppenheimer:2012} (solid). Note that these results are not directly comparable to the observational data 
from \protect\cite{DOdorico:2010}, as the spectra are not sampled to match the redshift distribution of that survey. We show the redshift-sampled WARM model (dotted) 
to quantify the size of this effect.
}
\label{fig:civ_compare}
\end{figure}

Figure~\ref{fig:civ_compare} shows the \civ~column density function from the $z=2$ output of our WARM simulation, compared to the results of \cite{Oppenheimer:2012} and the EAGLE simulation \protect\citep{Rahmati:2015}. 
Note that each simulation group computes the column density function slightly differently. Our methods are described above. \cite{Oppenheimer:2012} generated fake spectra in a similar way to us, but computed the column density 
by Voigt profile fitting. \cite{Rahmati:2015} instead projected the simulated \civ~density in the simulation box onto a cuboid grid, and computed the column density in each grid cell of 
volume~$(1 \mathrm{kpc})^2 \times 200 $~km/s. We do not expect these differences to be significant compared to the different feedback and simulation methods employed. All simulations produced reasonably similar results. 
In particular, WARM and EAGLE agree remarkably well at low column densities, $N_\mathrm{CIV} < 14$~\NHunit. Regrettably both produce fewer absorbers than are observed at these column densities. Interestingly, EAGLE produces 
substantially fewer strong absorbers than WARM, which agrees well with the data in this column density range. \cite{Oppenheimer:2012} produce the most absorbers with $N_\mathrm{CIV} < 14$~\NHunit, and is the only model to slightly over-produce the observed \civ~CDDF.
As explained in their paper, their metallicity-dependent wind model is very effective at enriching the CGM. Note that both simulations produce substantially fewer $N_\mathrm{CIV} > 15$~\NHunit absorbers than our WARM 
model; thus we expect both simulations to produce far fewer of the strong absorbers examined in Section \ref{sec:civglstrong}. The difference between the CDDF sampled to match the redshift distribution 
of~\cite{DOdorico:2010}~and the $z=2$ simulation output is comparable to the difference between EAGLE and WARM.

\subsection[Equivalent Widths of strong CIV]{Equivalent Width Distribution of Strong \civ~Absorbers}
\label{sec:civglstrong}

\begin{figure*}
\includegraphics[width=0.45\textwidth]{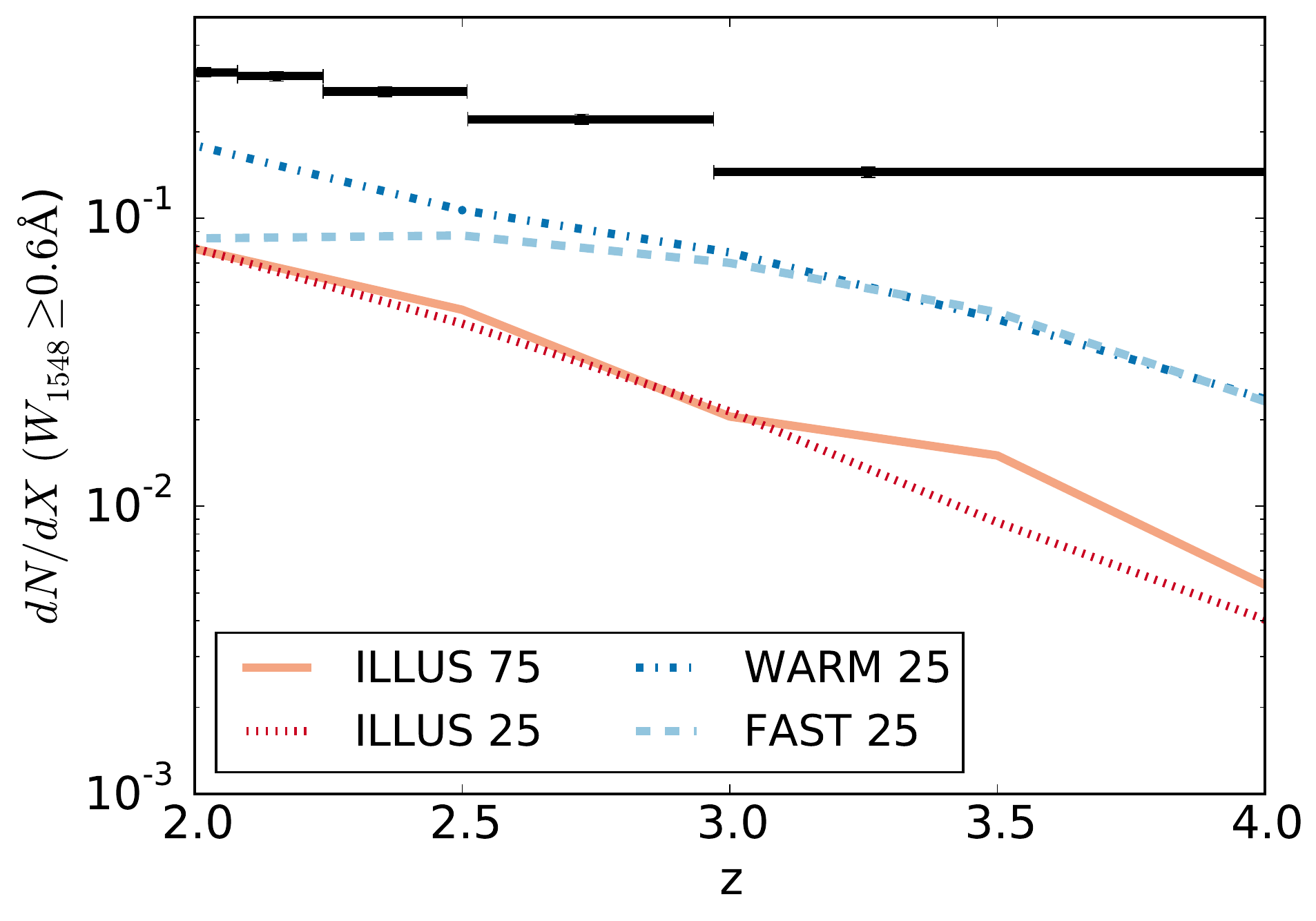}
\includegraphics[width=0.45\textwidth]{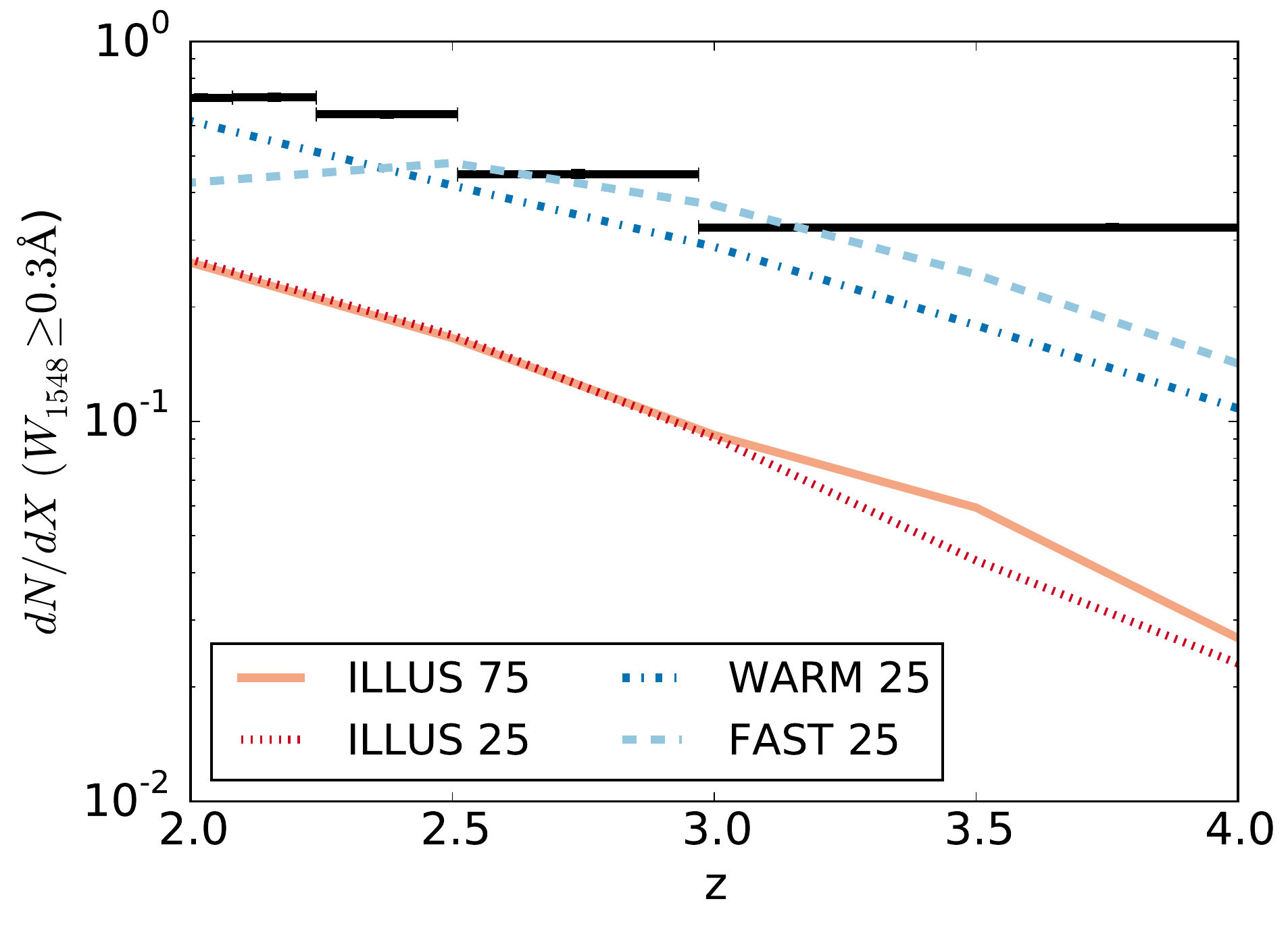} \\
\caption{The line density of \civ~extracted from our simulations at two different equivalent width thresholds, compared to the results 
of \protect\cite{Cooksey:2013}. Left: with $W_{1548} > 0.6 $~\AA. Right: with $W_{1548} > 0.3 $~\AA. 
Statistical error bars are included for the observations. \protect\cite{Cooksey:2013} may also systematically underestimate the true $dN/dX$ for $W_{1548} > 0.3 $~\AA~by $\sim 10\%$, 
due to the difficulty of detecting weak absorbers in SDSS spectra (Cooksey, private communication). 
Both statistical and systematic errors are small compared to the differences induced by our feedback models.
}
\label{fig:line_dens}
\end{figure*}

\begin{figure*}
\includegraphics[width=0.45\textwidth]{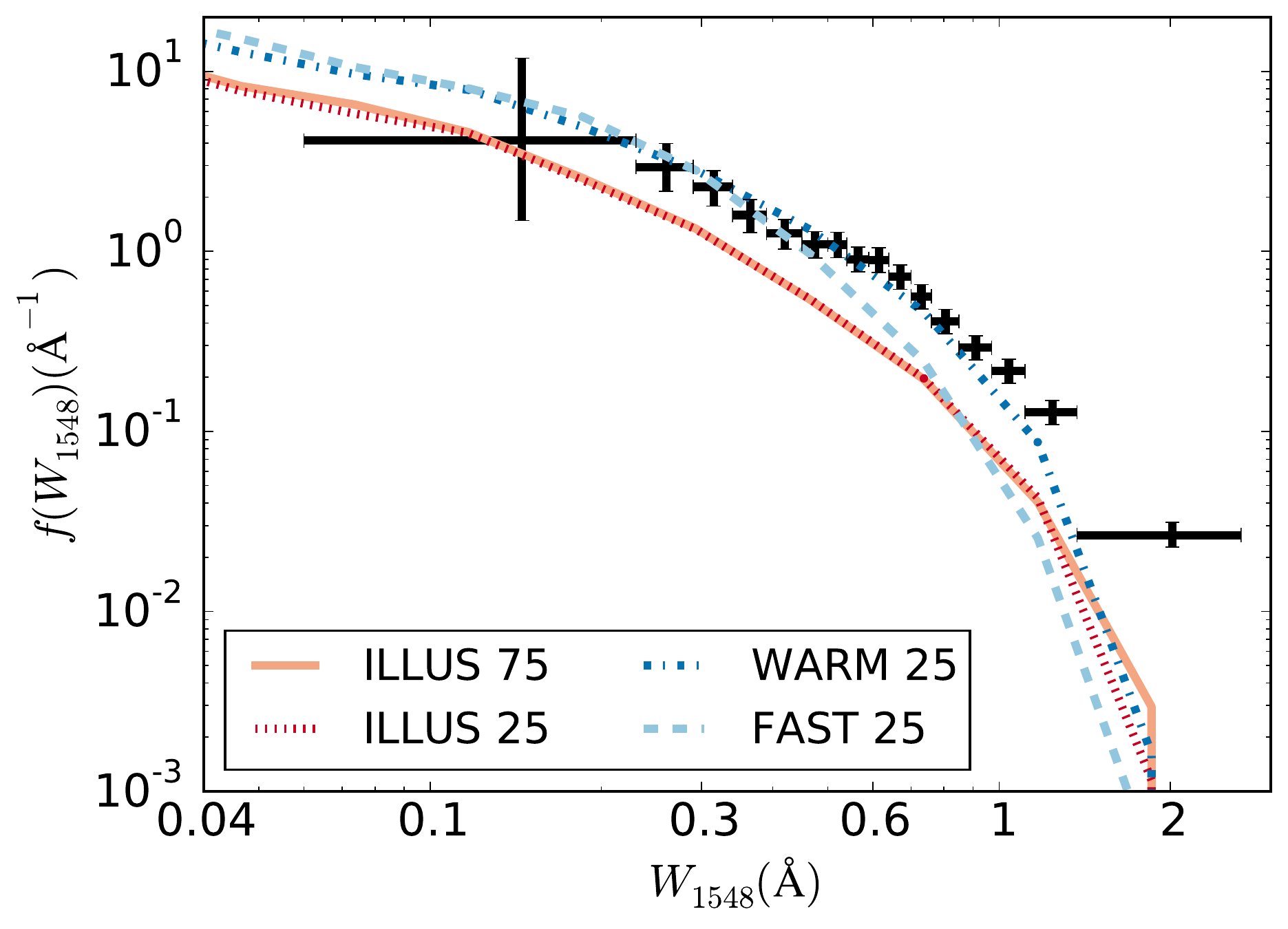}
\includegraphics[width=0.45\textwidth]{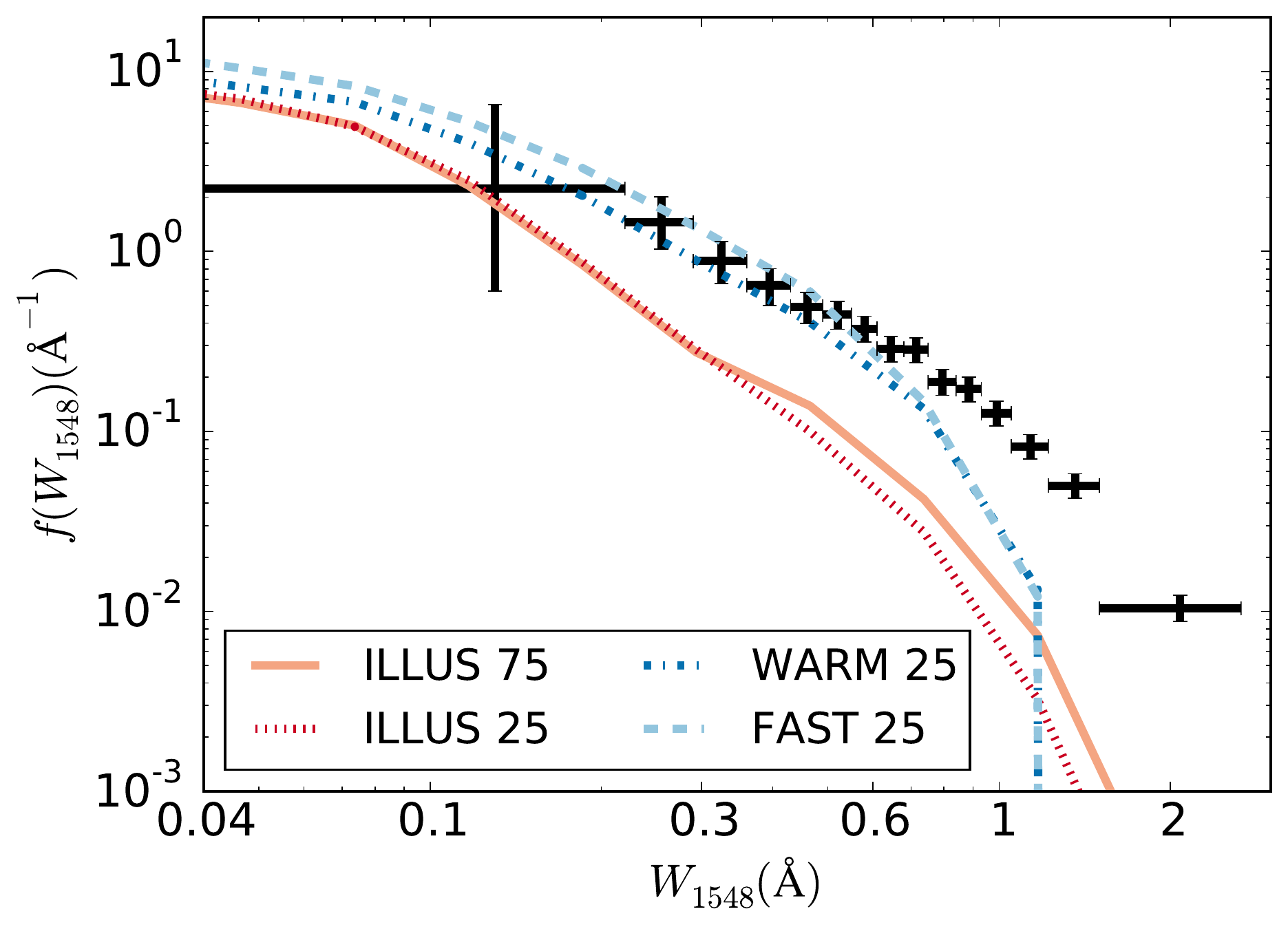}
\caption{The equivalent width distribution of \civ~extracted from our simulations, compared to the results of \protect\cite{Cooksey:2013}. 
Left: at $z=2$. Right: At $z=3.5$. As with the line density, warmer or faster winds are in substantially better agreement with observations than Illustris.
Note that sightlines with $N_\mathrm{CIV} = 10^{15}$ \NHunit, the upper limit used in Figure \protect\ref{fig:omega_civ}, have a median equivalent 
width of $W_{1548} \approx 0.5$ \AA~in our simulations.
}
\label{fig:eqw}
\end{figure*}

\cite{Cooksey:2013} detected and characterised a large sample of \civ~absorbers in the redshift range $z=2-4$ using SDSS spectra. 
The size of their catalogue allows them to constrain the statistics of strong absorbers. However, each individual SDSS spectrum 
is of relatively low quality, and so this survey suffers from incompleteness for absorbers with equivalent widths $W \lesssim 0.3$ \AA.
We compared this data to the same simulated catalogue as in Section \ref{sec:civglweak}; $10,000$ synthetic sightlines, with 
a \civ~column density greater than $10^{12}$~\NHunit, positioned at random through each simulation snapshot. 
We checked that sample variance was negligible for our simulated sample.

We define the \civ~equivalent width, $W_{1548}$, as the integral of the normalized flux along the sightline
\begin{equation}
W_{1548}(\tau ) = \int 1 - \exp(-\tau) dv\,.
\label{eq:eqwidth}
\end{equation}
The integration range was initially $-400 < v < 250 $~\kms~from the center of the absorber, which was defined as the velocity of the largest optical depth along the sightline.
The upper limit is set by the position of the \civ~$1550$\AA~line. The lower limit is extended by $100$ \kms if $\tau > 0.1$ anywhere in the 
extension region; in practice the largest window for an absorber was $-700 < v < 250$~\kms. This procedure follows \cite{Cooksey:2013} as closely as possible. We also considered
a fixed velocity interval of $-500 < v < 250$~\kms, with negligible change to our results.

Figure \ref{fig:line_dens} compares our simulated spectra to \cite{Cooksey:2013}. We show the observed line 
density per unit absorption distance, $\mathrm{d}N/\mathrm{d}X$, where $X$ is the absorption distance defined in Eq. \ref{eq:absdist}.
We show $\mathrm{d}N/\mathrm{d}X$ for absorbers with $W_{1548} > 0.3$ and $W_{1548} > 0.6$ \AA.
Sightlines with a \civ~equivalent width of $0.3$ \AA~have a median $N_\mathrm{CIV} \sim 5\times 10^{14}$~\NHunit~(integrated along the entire sightline)
in our simulations.

For both thresholds, the Illustris simulation produces less \civ~than is observed by a factor of $3$ at $z=2$ and a factor of $5$ at $z=3.5$.
The smaller and larger Illustris boxes, ILLUS 75 and ILLUS 25, which have identical feedback models, give similar results for $z<3$, indicating 
good convergence with box size at these redshifts. At higher redshifts the smaller box produces noticeably fewer \civ~absorbers. As discussed in Section~\ref{sec:absorbers}, these 
strong absorbers arise close to haloes which are relatively large for this redshift ($10^{11} - 10^{12} \Msun$), and the smaller box does not contain 
enough volume to sample this halo mass range completely.

By contrast, WARM and FAST, which employ more energetic wind models, each provide a good match to the observed $dN/dX(W> 0.3)$.
The increased enrichment is more pronounced at higher redshifts, where a larger proportion of star formation is found in small haloes. These haloes have 
shallower potential wells, allowing more expulsion of metals. Thus FAST and WARM also better match the observed redshift evolution of $dN/dX(W> 0.6)$.
However, even our more energetic wind models produce too few systems with an equivalent width greater than $0.6$ \AA.

At $z < 2.5$, FAST has a lower $dN/dX(W_{1548} > 0.6)$ than WARM. This is due to the inclusion of AGN feedback in FAST and ILLUS, but not WARM. 
AGN feedback heats the gas in large halos above the $10^5$ K where it can form \civ. 
We checked the effect of AGN feedback explicitly using two otherwise identical 
simulations\footnote{HVEL and HVNOAGN from \cite{Bird:2014}}, and confirmed that it can substantially suppress \civ~production 
at $z < 2.5$ and $W_{1548} > 0.3$ \AA. Note that FAST has a larger $dN/dX(W_{1548} > 0.3)$ than WARM at $z > 2.5$.

Figure \ref{fig:eqw} shows the equivalent width distributions from the simulations, compared to the measurements of \cite{Cooksey:2013}. 
We define the equivalent width distribution, $f(W_{1548})$, analogously to the column density distribution, as the number of 
absorbers in a given equivalent width bin, $N(W_{1548})$, per unit equivalent width per unit absorption distance, $\Delta X$ (Eq.~\ref{eq:absdist}):
\begin{align}
 f(W_{1548}) &= \frac{N(W_{1548})}{\Delta W_{1548} \Delta X(z)}\,.
\end{align}
We show the equivalent width distribution at two redshifts: $z=2$ and $z=3.5$. 
WARM and FAST both agree with the observations for $W_{1548} < 0.6 $~\AA~at $z=3.5$ and $W_{1548} < 0.3 $~\AA~at $z=2$. At $z=2$ agreement between WARM and the 
observations is good except for the strongest absorbers, but all other simulations have too few systems by an order of magnitude. The agreement between
WARM and the shape of the equivalent width distribution is remarkable, especially given that it lacks AGN feedback. 
At $z=3.5$ the WARM and FAST models, while similar, are not a good match to observations for $W_{1548} > 0.6 $~\AA.
The similarity between WARM and FAST indicates that the lack of AGN feedback in WARM is not having a significant effect at $z > 3$.
At $z=3.5$ the $25$ Mpc/h simulations have reduced numbers of strong \civ absorbers due to their small simulation volume. However, 
the discrepancy between WARM and the observations is larger than the difference between the $25$ and $75$ Mpc/h boxes, indicating that a larger WARM 
box is unlikely to fully explain the discrepancy with observations. 

Overall, both $dN/dX$ and $f(W_{1548})$ paint a similar picture; good agreement between WARM and the observations for absorbers with $W_{1548} < 0.3 $~\AA, but a 
lack of the strongest absorbers, especially at $z=3.5$. Interestingly, we showed in Figure~\ref{fig:civ_compare} that other modern simulations, EAGLE and 
that of \cite{Oppenheimer:2012}, produce slightly fewer strong absorbers than WARM. Given this, we expect that both simulations would also 
under-produce \civ~absorbers with equivalent widths $W_{1548} > 0.6$~\AA, as we do, suggesting that
the lack of large equivalent width absorbers in our simulations is a generic problem with modern feedback models.

Most absorbers with $W > 0.6$~\AA~are found around haloes with masses $10^{11} - 10^{12} \Msun$ (see Figure \ref{fig:civhalomass}).
A common feature of all current supernova feedback models is that they become ineffective at suppressing star formation 
in high mass haloes  \citep[e.g.~][]{Vogelsberger:2013}, motivating the need for AGN feedback. The reason behind this is that 
the deeper potential wells of these halos render the supernova feedback models unable to expel gas from star-forming regions. 
It seems that matching the abundance of $W_{1548} > 0.6$~\AA~absorbers also requires additional gas to be expelled 
from star-forming regions in large halos. This could perhaps be achieved by stellar feedback models which are still efficient 
in larger halos, or by modifying the AGN feedback to expel gas without heating it beyond $10^5$K.

\subsection[Carbon near DLAs]{Carbon absorption near Damped Lyman-$\alpha$ Systems}
\label{sec:civdlas}

\begin{figure*}
\includegraphics[width=0.45\textwidth]{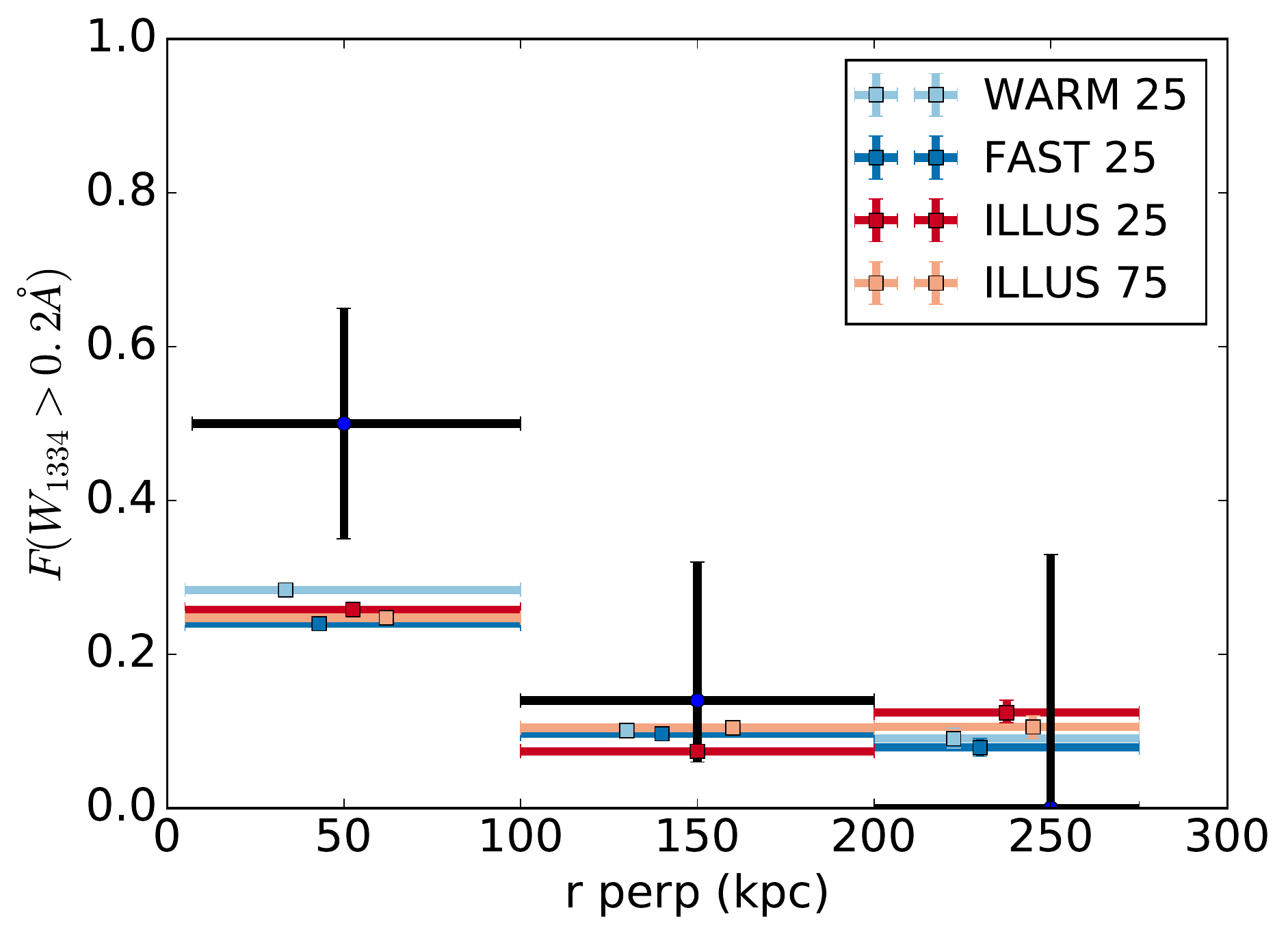}
\includegraphics[width=0.45\textwidth]{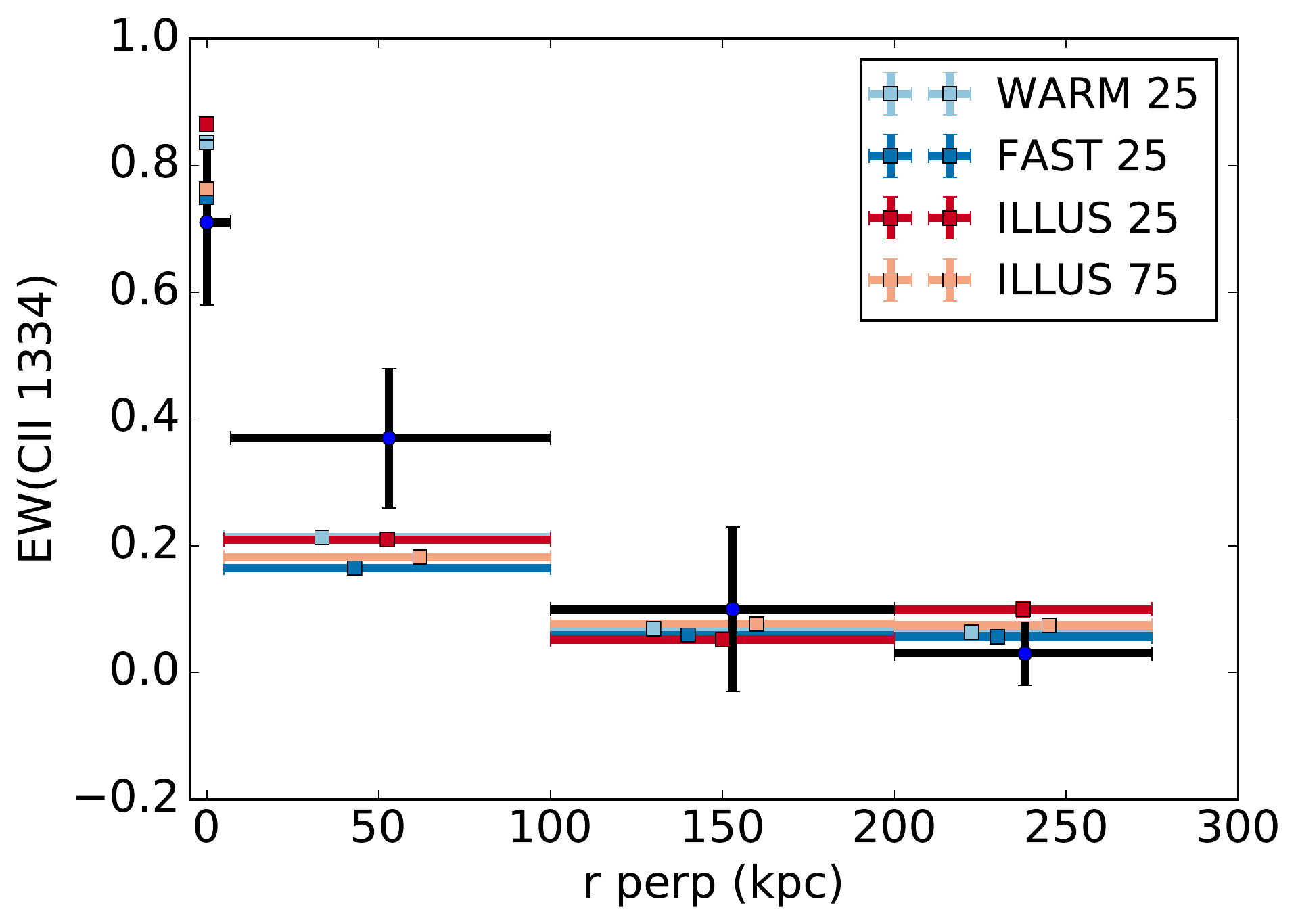}
\caption{Covering fractions (Left) and mean equivalent widths (Right) of \cii~in $100$ kpc wide impact parameter bins around DLAs, 
compared to the observations of \protect\cite{Rubin:2014} (black error bars).
Bin centroids are offset for clarity. Vertical error bars on the simulations are generated using bootstrap resampling and show that 
the effect of sample variance is negligible. Points at $r_\mathrm{perp} =0$ indicate the equivalent width at the DLA.
}
\label{fig:ciieqw}
\end{figure*}

\begin{figure*}
\includegraphics[width=0.45\textwidth]{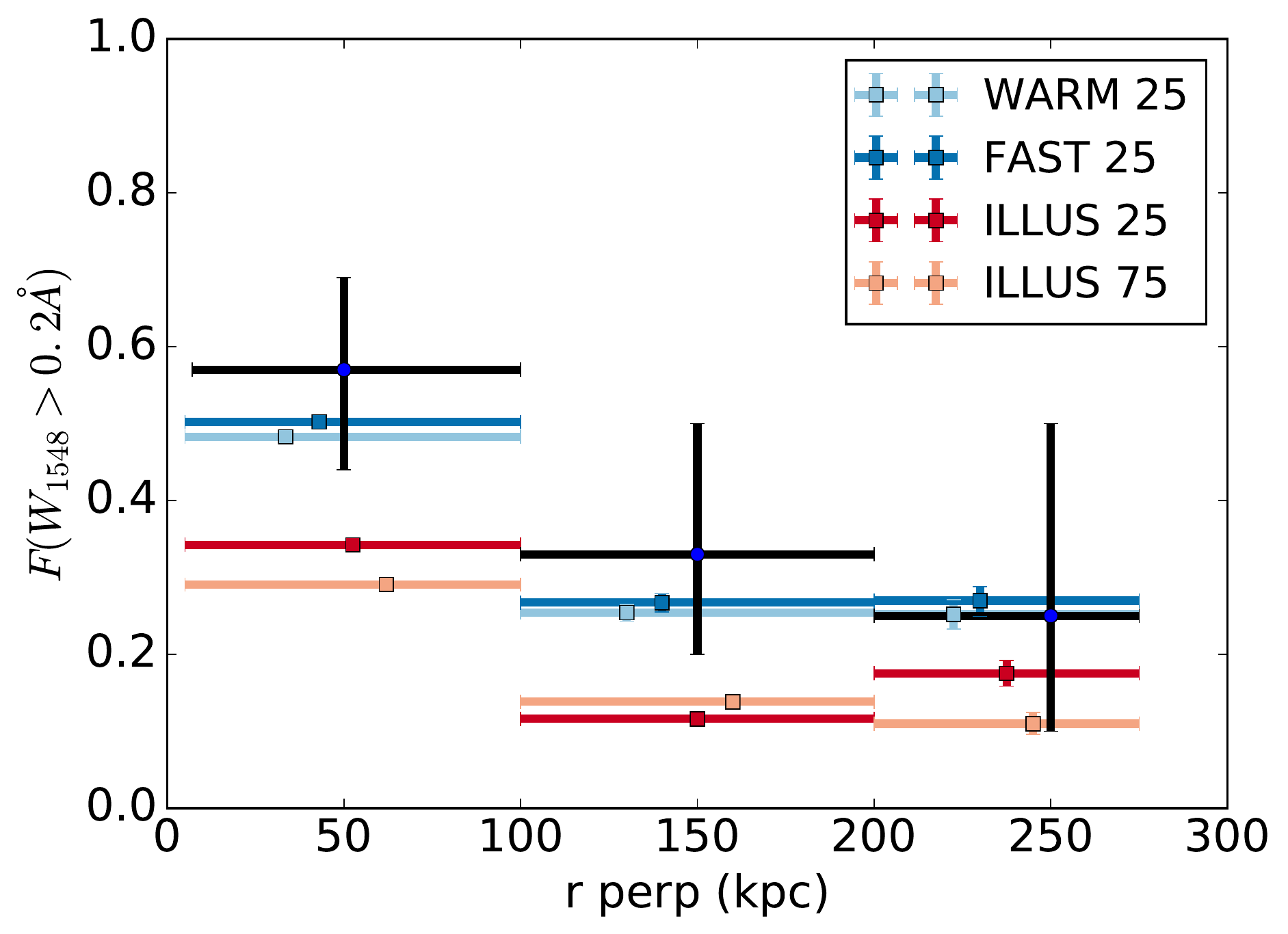}
\includegraphics[width=0.45\textwidth]{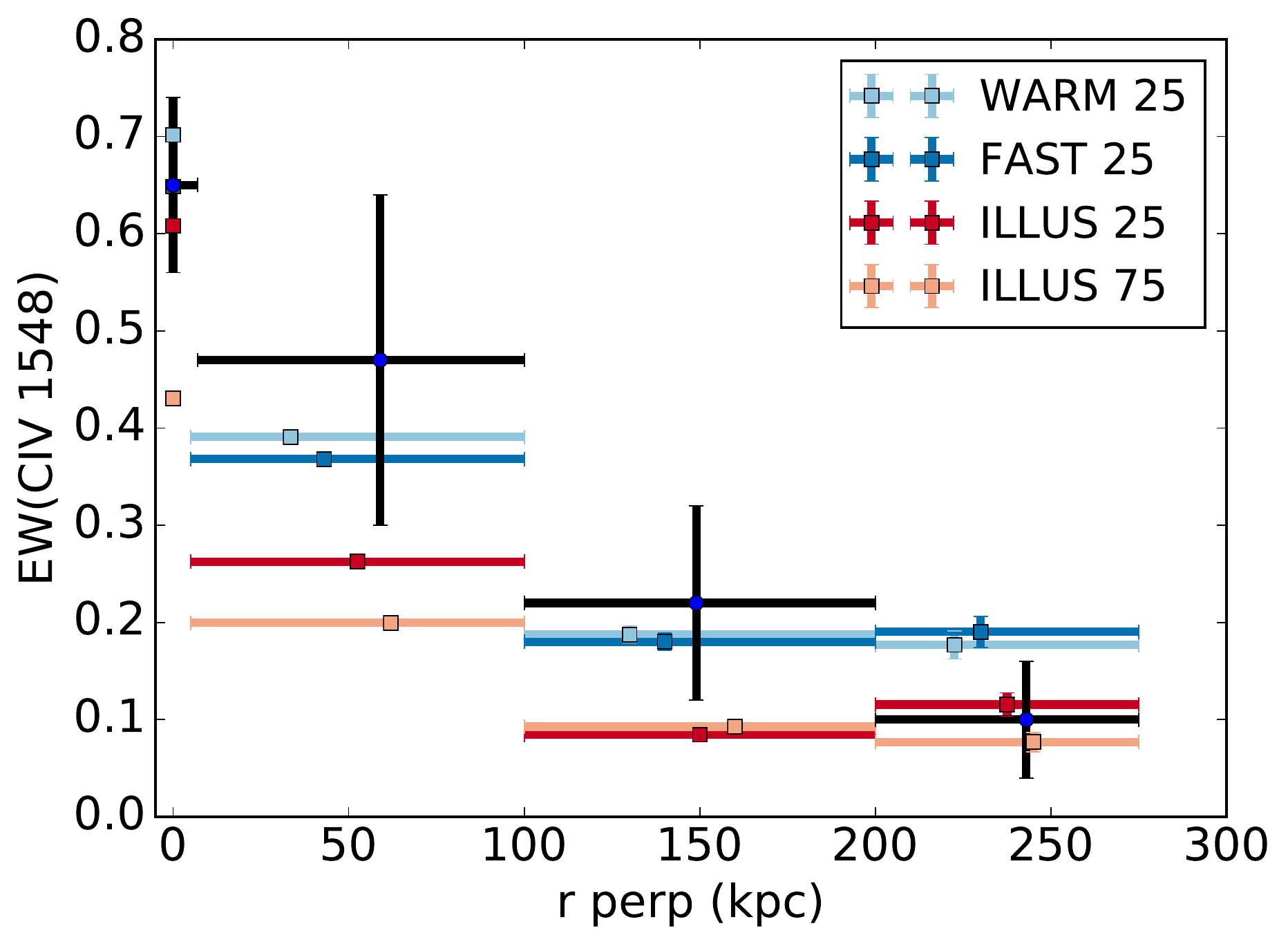}
\caption{Covering fractions (Left) and mean equivalent widths (Right) of \civ~in $100$ kpc wide impact parameter bins around DLAs, 
compared to the observations of \protect\cite{Rubin:2014} (black error bars).
Bin centroids are offset for clarity. Vertical error bars on the simulations are generated using bootstrap resampling and show that 
the effect of sample variance is negligible. Points at $r_\mathrm{perp} =0$ indicate the equivalent width at the DLA. 
}
\label{fig:civcover}
\end{figure*}



The equivalent widths of \civ~and \cii~around DLAs were recently measured by \cite{Rubin:2014}. 
We have shown above that our simulations differ in the extent to which they expel metals into the circumgalactic 
medium, and DLAs are a tracer of the low-mass galaxy population at $z=2-4$. Comparisons with the enrichment around DLAs thus seems 
a promising way to observationally constrain feedback models. It further provides a test of our conclusions in 
Sections \ref{sec:civglweak} and \ref{sec:civglstrong} by partially isolating the effects on smaller haloes.
All our feedback models produce similar populations of DLAs, in good agreement with observations \citep{Bird:2014, Bird:2014a}.
In particular, they match the observed DLA metallicity distribution and produce DLA host halos with realistic virial velocities. 
We find that the majority of DLAs are found within haloes of mass $5\times 10^{9} - 10^{11} \Msun$, and virial velocities $50-150$~\kms.

\cite{Rubin:2014} obtained spectra from $40$ quasar pairs and lensed quasars with sightlines separated by $7.5$ to $275$ kpc (physical). These pairs were selected 
so that a DLA was present in one sightline of the pair. Measurements of absorbers at the redshift of the DLA in the non-DLA sightline 
can thus constrain the state of the gas surrounding the DLA. As in previous sections, we generated a synthetic sightline catalogue similar to that observed. 
We first selected sightlines containing DLAs. The DLAs observed by \cite{Rubin:2014} are concentrated at 
redshifts $z=2-3$, so we selected sightlines from our snapshots at $z=2$, $2.5$ and $3$ weighted to match the observed redshift distribution.
For each of these DLAs we chose a sightline with a random impact parameter. The distribution of impact parameters was constrained to match 
the observed sample in each redshift bin, although in practice this was less important than matching the observed redshift 
distribution. Our simulated sample contained a total of $14,000$ quasar pairs containing a DLA.

We computed equivalent widths following the method of \cite{Rubin:2014} as far as possible. We found the strongest HI absorption 
in the CGM sightline within $\pm 600$ \kms~of the DLA redshift. \cite{Rubin:2014}~estimated, by eye, the velocity extent of this single HI absorption 
component and looked within this region for a strong metal line. We mimic this by choosing a region initially $400$ \kms across, the mean size 
of the by-eye search window chosen in the observations, and extending it as needed in $100$ \kms increments until we encounter a region without 
significant \Lya~absorption, defined as a mean HI optical depth $\tau_\mathrm{HI} > 0.1$ in the new $100$ \kms region. We then examine this region to find
the strongest metal line and defined the equivalent width to be the integral within $\pm 300$ \kms~around 
line centre. Note that this procedure gave results comparable to defining the equivalent width to be the integral of the metal line 
absorption within $\pm 600$ \kms of the DLA.

Figure \ref{fig:ciieqw} shows the covering fractions of \cii~systems with $W_{1334} > 0.2$~\AA, $F(W_{1334}>0.2)$, and the mean \cii~equivalent widths, 
for sightlines around DLAs. 
\cii~traces cooler, denser gas than \civ, more closely associated with a DLA or a Lyman Limit System.
This cold gas is insensitive the differences between our feedback models; the outflowing gas shock-heats and 
so yields negligible \cii~absorption in any case.
Thus all our simulations give very similar results, in good agreement with the observational measurements 
at $r_\mathrm{perp} > 100$ kpc, but $1-2\sigma$ discrepant at $r_\mathrm{perp} < 100$ kpc.
The good agreement of the simulations at larger perpendicular distances is another useful validation of our underlying model.
The moderately low \cii~equivalent width nearer the DLA predicted in our models, combined with the overly large value of $\Omega_\mathrm{DLA}$ at $z=2$ noted in \cite{Bird:2014}, 
may suggest that the feedback model needs to further suppress cold gas in small haloes. This would cause the median DLA to reside in a slightly 
more massive halo, slightly boosting its metallicity, and thus \cii~abundance. However, the significance of the discrepancy is not high, and it 
could simply be a statistical fluctuation, especially as the models are in good agreement with the observed \cii~equivalent width within the DLAs. We have also 
checked that all our simulations give similar results for the observed covering fractions of \SiII~and \SiIV~around DLAs, in good agreement 
with the observations of~\cite{Rubin:2014}.

Figure \ref{fig:civcover} shows the covering fractions of \civ~systems with $W_{1548} > 0.2$~\AA, $F(W_{1548}>0.2)$, and the mean \civ~equivalent widths, 
for sightlines around DLAs. The \civ~statistics are sensitive to the differences between our feedback models, 
and the more energetic winds in WARM and FAST are again in better agreement with observations than the default Illustris model, especially near the DLA. 
We have checked that the increased \civ~equivalent width owes to a larger column density, rather than increased thermal or velocity broadening in existing lines.
In the bin farthest from the DLA, the WARM simulation slightly overpredicts the mean \civ~equivalent width, but we do not regard this as significant, 
especially as the \civ~covering fraction in the same bin is in good agreement with observations. The tension between the Illustris feedback model and the 
observed \civ~statistics is relatively weak, but certainly significant, especially when considered with the results of earlier sections. 
The statistics of \civ~around DLAs thus also favour the WARM and FAST feedback models, and extend our conclusions by isolating the effect of 
the feedback model on the smaller halos which host DLAs.

\section{Conclusions}
\label{sec:conclusions}

We have compared our simulations to three different carbon absorber measurements at $z=2-4$, with the aim of constraining a parameter of the Illustris feedback model. 
Two are blind \civ~surveys; the column density distribution function (CDDF) of \civ~absorbers from \cite{DOdorico:2010} and the equivalent width distribution of 
(stronger) \civ~absorbers from \cite{Cooksey:2013}. The third measurement is a \civ~and~\cii~absorption survey around DLAs from \cite{Rubin:2014}. 
We compared to the Illustris simulation and some smaller simulations with modified feedback models. Motivated by the results of \cite{Suresh:2015}, 
our feedback models were modified to have more energetic winds, realised by an increased wind velocity or an increased wind thermal energy.
Our modifications to the Illustris feedback model were chosen so that they did not alter the good agreement with the observed galaxy stellar mass function.

We demonstrated that, nevertheless, our more energetic feedback models affected the CDDF and equivalent width 
distributions of \civ~at a level highly significant compared to the observational error bars, and the distribution of \civ~around DLAs at a level 
comparable to the error bars. All measurements favoured the more energetic wind models, which were better able to enrich the gas 
surrounding haloes. There is little difference between the model with an increased wind velocity and that with increased thermal energy, 
so we do not favour one over the other, both being in generally good agreement with observations. The Illustris feedback model did not 
sufficiently enrich the gas surrounding galaxies, and under-produced \civ~in all the observational measurements we compared to.
Furthermore, \cite{Suresh:2015}~found that a larger wind energy per unit mass moderately improves agreement 
with the radial profile of \civ~around Lyman break galaxies (LBGs) \citep{Turner:2014}. All wind models produce similar results for 
the \cii~around DLAs, which is in reasonable good agreement with observations, being discrepant at the $1.5-\sigma$ level 
in only one projected radial bin near the DLA.

We were not able to reproduce the abundance of the strongest \civ~absorbers with equivalent widths $W_{1548} > 0.6$~\AA.
The SDSS spectra within which these absorbers are detected are low resolution and of relatively low signal to noise.
They may thus suffer from various observational systematics, including blending with other metal lines or poor continuum estimation.
However, to completely reconcile our results, these effects would have to be large; we under-produce $W_{1548} \sim 1 $~\AA~absorbers 
by almost an order of magnitude.

Our simulations thus still appear to suffer from a generic lack of \civ~absorbers with the highest column densities. It may be possible to resolve this discrepancy 
by further increasing the energy per unit mass of the wind model. To avoid over-producing lower column densities or reducing the stellar mass function below
the range allowed by observations, such a change would need to be targeted at $10^{11}-10^{12} \Msun$ halos, a mass range where supernova winds struggle 
to escape their halos. Alternatively, since these strong absorbers are saturated,
it may be possible to achieve better agreement with observations by increasing the velocity dispersion of the absorbing gas rather than the \civ~column density;
in this case it would be necessary to avoid heating the gas substantially above $10^5$ K and further ionizing carbon.
A final possibility could be a modification of the AGN feedback model to enrich the gas without heating it beyond $10^5$~K.

\cite{Rahmati:2015} measured the \civ~column density function in the EAGLE Project, a recent simulation using purely thermal supernova feedback 
models. We performed a comparison of the results of the EAGLE project, as well as the simulations of \cite{Oppenheimer:2012}, to ours.
While both simulations, \cite{Oppenheimer:2012} especially, were a reasonable match to the \civ~CDDF, they both produced fewer strong absorbers than our most energetic 
wind models. Given this, we expect that both simulations would also under-produce \civ~absorbers with equivalent widths $W_{1548} > 0.6$~\AA, as we do, suggesting that
the lack of large equivalent width absorbers in our simulations is a generic problem with modern feedback models.

We examined some properties of the \civ~absorbers, including their host halos, their distribution within a halo, and their collisional ionization fraction.
We found that \civ~absorbers are found throughout halos, even substantially beyond their virial radius, while \cii~absorbers are generally confined to the 
high density gas near the inner regions of the halo. The strongest \civ~absorbers with $W_{1548} > 0.5$~\AA~are however associated with the interiors of 
the larger haloes. Especially in the simulations with stronger winds, \civ~absorbers of all strengths arise predominantly from photoionized gas at $z=2-3$. 

This paper should be viewed in the context of our previous work \citep{Bird:2014}, which showed that the Illustris simulation reproduces the statistics of 
strong neutral hydrogen absorbers. We showed there that the fast and warm winds considered in this paper produce a very similar population of neutral hydrogen 
absorbers to the default Illustris model, and we showed here that they do not affect the galaxy stellar mass function. This thus demonstrates that \civ~absorbers 
are able to constrain the Illustris feedback model parameters in a way independent to either of these quantities.
Taken together, these papers may indicate that it is possible to calibrate a feedback model to high redshift absorption 
line statistics, rather than the more usual low redshift stellar mass function. Absorption lines can probe low-mass objects at high redshift whose stellar components 
are not directly observable. Furthermore, they have very different systematic errors from galaxy stellar mass functions. We may return to this in future work.

Overall, we recommend that designers of future supernova feedback models explicitly check whether they match the observed properties 
of \civ~in the intergalactic medium or around galaxies. Models which include winds with sufficient energy to allow metals to easily escape galactic haloes 
and enrich their environment seem likely to provide a reasonable match to the \civ~CDDF, as shown both in our results and those of \cite{Oppenheimer:2012}.
While the Illustris model is somewhat discrepant, we showed that our warm and fast wind models produce gas around relatively low-mass haloes in 
good agreement with the \civ~CDDF, and the distribution of \civ~and~\cii~around DLAs. However, they under-produce the number of \civ~absorbers 
with $W_{1548} > 0.6$~\AA, necessitating some further modification of the feedback model.

\appendix

\section{Resolution Convergence}
\label{sec:resolution}

\begin{figure}
\includegraphics[width=0.45\textwidth]{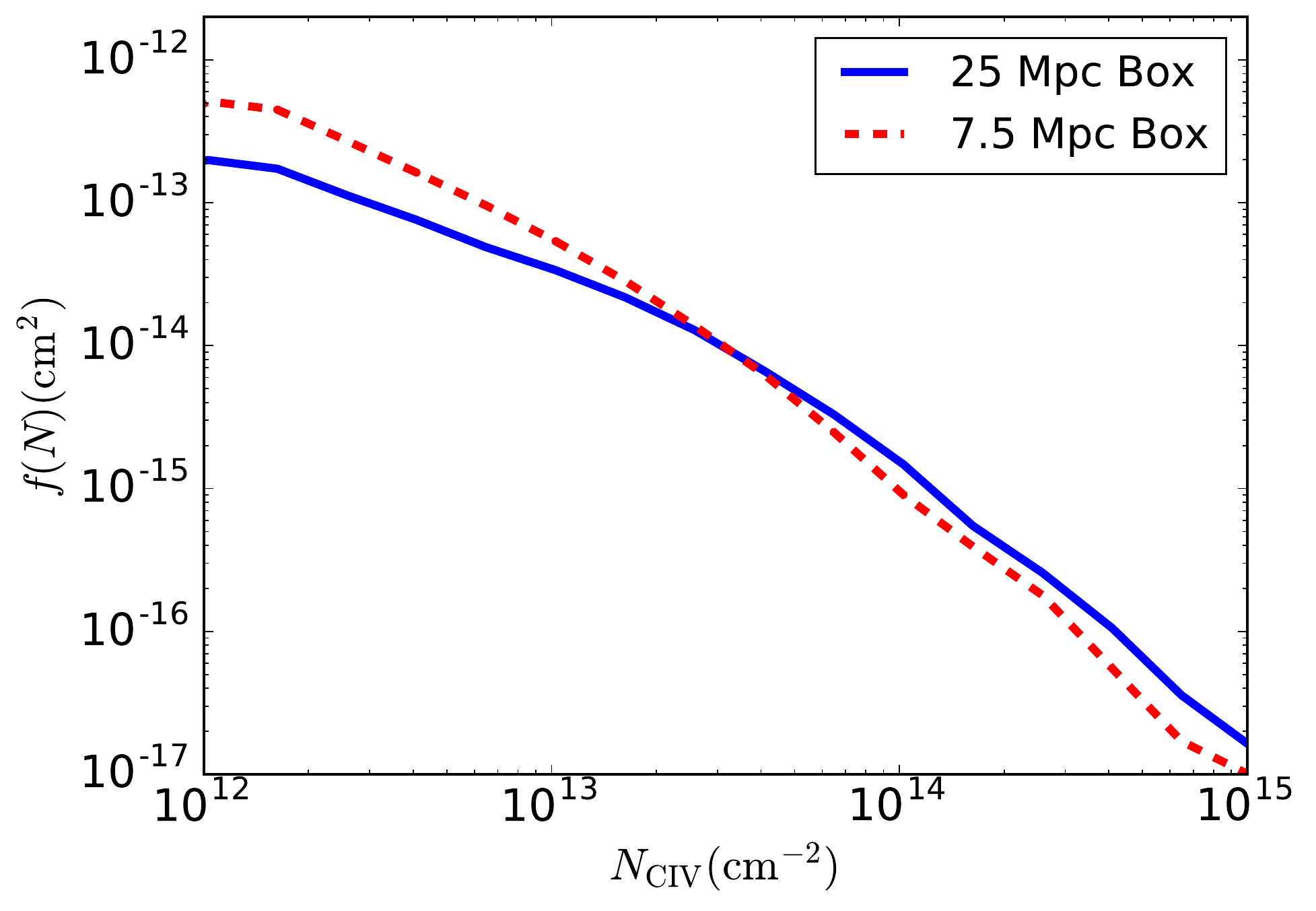}
\caption{A comparison of the \civ~column density distribution from our $25$ Mpc Illustris feedback model simulation and an identical $7.5$ Mpc box with higher resolution.
To limit the effect of the smaller box, we have removed from the column density function spectra near halos with $M > 2.7 \times 10^{11} M_\odot/h$, the mass of the largest halo in the smaller box.
We caution that for  $N_\mathrm{CIV} > 10^{14}$ \NHunit, over $90\%$ of the absorbers in the larger box are associated with halos with mass $M > 10^{11} M_\odot/h$, of which there are only $8$ in the smaller box.
}
\label{fig:resolution}
\end{figure}

In this Appendix, we will consider how our results are affected by the resolution of the simulation. Computational limits prevent us from performing a simulation with the same box size 
and double the resolution of our fiducial box. Instead we have performed a simulation which uses the Illustris feedback model, but reduces the box size to $7.5$ Mpc.
The particle load remains constant at $512^3$, thus increasing the mass resolution by a factor of $\sim 40$. Unfortunately, the $7.5$ Mpc box artificially suppresses the number of absorbers with 
$N_\mathrm{CIV} > 10^{14}$ \NHunit because it is insufficiently large to include many of the more massive halos. The largest halo in the $7.5$ Mpc/h box has mass $M = 2.7 \times 10^{11} M_\odot/h$, while the
largest halo in the $25$ Mpc/h box has $M = 2\times 10^{12} M_\odot/h$. To limit the effect of these larger halos, we exclude from the sample all spectra from the $25$ Mpc box associated with halos 
with mass $M = 2.7 \times 10^{11} M_\odot/h$. Figure \ref{fig:resolution} shows the resulting \civ~CDDF for both boxes. The small box reduces the number of \civ~absorbers with $N_\mathrm{CIV} > 10^{14}$ \NHunit
by a factor of two. This difference is most likely due to the reduced box size; over $90\%$ of the CDDF in the $25$ Mpc/h box comes from absorbers associated with halos with mass $M > 10^{11} M_\odot/h$. 
This mass range includes only $8$ halos in the $7.5$ Mpc/h box, and so the sample size is still insufficient.

The increase in resolution in the $7.5$ Mpc/h box increases the number of absorbers with $N_\mathrm{CIV} < 10^{13}$ \NHunit by up to $50\%$. This arises from the increase in the global star formation rate from resolving smaller halos. 
The smaller box resolves halos down to $10^9 M_\odot/h$, and thus includes essentially all halos able to form stars against the thermal pressure of the ionized intergalactic medium. Compared to the differences between 
our Illustris and WARM feedback models, the increase with higher resolution is relatively mild. It is interesting to note that a correction of this size, applied to the WARM model, would substantially improve agreement with observations.

We do not show resolution convergence for the equivalent width distribution shown in Section \ref{sec:civglstrong}, as these strong absorbers are found almost exclusively in halos more massive than those in the $7.5$ Mpc box.

\section{Tabulated Results}
\label{sec:tables}

In this Appendix we provide our results in tabulated form. Table \ref{tab:civ_cddf_4_25.txt} gives the CIV CDDF from Figure \ref{fig:omega_civ}, 
and Table \ref{tab:om_civ_9_25.txt} gives the evolution of $\Omega_\mathrm{CIV}$. Tables \ref{tab:eq_width_9_25_2.txt} and \ref{tab:eq_width_9_25_5.txt}
give the equivalent width distributions of strong absorbers from Figure \ref{fig:eqw} at $z=3.5$ and $z=2$ respectively. Table \ref{tab:lciv_06_4_25.txt} 
shows Figure \ref{fig:line_dens}, the evolution of dN/dX ( $W_{1548} > 0.6$\,\AA) with redshift.

\begin{table} 
 \centering 
\begin{tabular}{cccc}
\hline
$N_\mathrm{CIV}$ (cm$^{-2}$) & WARM  & FAST  & ILLUS 75  \\ 
 \hline 
$ 1.0 \times 10^{ 12 }$ & $ 5.6 \times 10^{ -13 }$ & $ 7.3 \times 10^{ -13 }$ & $ 4.1 \times 10^{ -13 }$  \\ 
$ 1.6 \times 10^{ 12 }$ & $ 4.3 \times 10^{ -13 }$ & $ 5.6 \times 10^{ -13 }$ & $ 2.9 \times 10^{ -13 }$  \\ 
$ 2.6 \times 10^{ 12 }$ & $ 2.6 \times 10^{ -13 }$ & $ 3.3 \times 10^{ -13 }$ & $ 1.7 \times 10^{ -13 }$  \\ 
$ 4.1 \times 10^{ 12 }$ & $ 1.7 \times 10^{ -13 }$ & $ 2.1 \times 10^{ -13 }$ & $ 1.1 \times 10^{ -13 }$  \\ 
$ 6.5 \times 10^{ 12 }$ & $ 1.0 \times 10^{ -13 }$ & $ 1.3 \times 10^{ -13 }$ & $ 7.2 \times 10^{ -14 }$  \\ 
$ 1.0 \times 10^{ 13 }$ & $ 6.7 \times 10^{ -14 }$ & $ 8.2 \times 10^{ -14 }$ & $ 4.4 \times 10^{ -14 }$  \\ 
$ 1.6 \times 10^{ 13 }$ & $ 4.0 \times 10^{ -14 }$ & $ 5.2 \times 10^{ -14 }$ & $ 2.8 \times 10^{ -14 }$  \\ 
$ 2.6 \times 10^{ 13 }$ & $ 2.6 \times 10^{ -14 }$ & $ 3.3 \times 10^{ -14 }$ & $ 1.7 \times 10^{ -14 }$  \\ 
$ 4.1 \times 10^{ 13 }$ & $ 1.6 \times 10^{ -14 }$ & $ 1.9 \times 10^{ -14 }$ & $ 9.5 \times 10^{ -15 }$  \\ 
$ 6.5 \times 10^{ 13 }$ & $ 8.9 \times 10^{ -15 }$ & $ 1.1 \times 10^{ -14 }$ & $ 5.2 \times 10^{ -15 }$  \\ 
$ 1.0 \times 10^{ 14 }$ & $ 5.1 \times 10^{ -15 }$ & $ 6.6 \times 10^{ -15 }$ & $ 2.4 \times 10^{ -15 }$  \\ 
$ 1.6 \times 10^{ 14 }$ & $ 2.9 \times 10^{ -15 }$ & $ 3.5 \times 10^{ -15 }$ & $ 1.2 \times 10^{ -15 }$  \\ 
$ 2.6 \times 10^{ 14 }$ & $ 1.5 \times 10^{ -15 }$ & $ 1.6 \times 10^{ -15 }$ & $ 5.3 \times 10^{ -16 }$  \\ 
$ 4.1 \times 10^{ 14 }$ & $ 7.4 \times 10^{ -16 }$ & $ 7.0 \times 10^{ -16 }$ & $ 2.6 \times 10^{ -16 }$  \\ 
$ 6.5 \times 10^{ 14 }$ & $ 3.8 \times 10^{ -16 }$ & $ 2.9 \times 10^{ -16 }$ & $ 1.2 \times 10^{ -16 }$  \\ 
$ 1.0 \times 10^{ 15 }$ & $ 1.6 \times 10^{ -16 }$ & $ 1.0 \times 10^{ -16 }$ & $ 6.3 \times 10^{ -17 }$  \\ 
\hline 
  \end{tabular}
 \caption{CIV CDDF $f(N)$ in cm$^2$ for our simulations.}
\label{tab:civ_cddf_4_25.txt}
 \end{table}

\begin{table} 
 \centering 
\begin{tabular}{cccc}
\hline
$z$ & FAST  & ILLUS 75  & WARM  \\ 
 \hline 
$ 4.00 $ & $ 1.79 $ & $ 0.47 $ & $ 1.33 $  \\ 
$ 3.50 $ & $ 2.74 $ & $ 0.78 $ & $ 2.00 $  \\ 
$ 3.00 $ & $ 4.01 $ & $ 1.12 $ & $ 3.02 $  \\ 
$ 2.50 $ & $ 5.36 $ & $ 1.85 $ & $ 4.29 $  \\ 
$ 2.00 $ & $ 5.31 $ & $ 2.61 $ & $ 6.07 $  \\ 
\hline 
  \end{tabular}
 \caption{$\Omega_\mathrm{CIV} \times 10^{8}$ for our simulations.}
\label{tab:om_civ_9_25.txt}
 \end{table}
 
\begin{table} 
 \centering 
\begin{tabular}{cccc}
\hline
$W_{1548}$ (\AA)  & FAST  & ILLUS 75  & WARM  \\ 
 \hline 
$ 4.7 \times 10^{ -2 }$ & $ 15.22 $ & $ 8.29 $ & $ 12.79 $  \\ 
$ 7.4 \times 10^{ -2 }$ & $ 10.58 $ & $ 6.54 $ & $ 9.67 $  \\ 
$ 0.12 $ & $ 8.02 $ & $ 4.58 $ & $ 7.89 $  \\ 
$ 0.19 $ & $ 5.61 $ & $ 2.56 $ & $ 4.91 $  \\ 
$ 0.29 $ & $ 2.81 $ & $ 1.33 $ & $ 2.82 $  \\ 
$ 0.47 $ & $ 0.97 $ & $ 0.54 $ & $ 1.32 $  \\ 
$ 0.74 $ & $ 0.24 $ & $ 0.20 $ & $ 0.46 $  \\ 
$ 1.17 $ & $ 2.6 \times 10^{ -2 }$ & $ 4.0 \times 10^{ -2 }$ & $ 8.7 \times 10^{ -2 }$  \\ 
$ 1.85 $ & $ 4.3 \times 10^{ -4 }$ & $ 3.0 \times 10^{ -3 }$ & $ 1.6 \times 10^{ -3 }$  \\ 
\hline 
  \end{tabular}
 \caption{Equivalent width distributions, $f(W_{1548})$ (\AA$^{-1}$), for our simulations at $z=2$.}
\label{tab:eq_width_9_25_5.txt}
 \end{table}

\begin{table} 
 \centering 
\begin{tabular}{cccc}
\hline
$W_{1548}$ (\AA)  & FAST  & WARM  & ILLUS 75  \\ 
 \hline 
$ 4.7 \times 10^{ -2 }$ & $ 10.40 $ & $ 8.20 $ & $ 6.68 $  \\ 
$ 7.4 \times 10^{ -2 }$ & $ 8.27 $ & $ 6.76 $ & $ 5.00 $  \\ 
$ 0.12 $ & $ 5.26 $ & $ 4.02 $ & $ 2.32 $  \\ 
$ 0.19 $ & $ 2.90 $ & $ 2.04 $ & $ 0.84 $  \\ 
$ 0.29 $ & $ 1.36 $ & $ 0.89 $ & $ 0.28 $  \\ 
$ 0.47 $ & $ 0.60 $ & $ 0.40 $ & $ 0.14 $  \\ 
$ 0.74 $ & $ 0.15 $ & $ 0.13 $ & $ 4.2 \times 10^{ -2 }$  \\ 
$ 1.17 $ & $ 1.2 \times 10^{ -2 }$ & $ 1.3 \times 10^{ -2 }$ & $ 7.4 \times 10^{ -3 }$  \\ 
$ 1.85 $ & $0$ & $0$ & $ 3.3 \times 10^{ -4 }$  \\ 
\hline 
  \end{tabular}
 \caption{Equivalent width distributions, $f(W_{1548})$ (\AA$^{-1}$), for our simulations at $z=3.5$.}
\label{tab:eq_width_9_25_2.txt}
 \end{table}

\begin{table} 
 \centering 
\begin{tabular}{cccc}
\hline
$z$ & WARM  & ILLUS 75  & FAST  \\ 
 \hline 
$ 4.00 $ & $ 2.4 \times 10^{ -2 }$ & $ 5.3 \times 10^{ -3 }$ & $ 2.3 \times 10^{ -2 }$  \\ 
$ 3.50 $ & $ 4.4 \times 10^{ -2 }$ & $ 1.5 \times 10^{ -2 }$ & $ 4.7 \times 10^{ -2 }$  \\ 
$ 3.00 $ & $ 7.6 \times 10^{ -2 }$ & $ 2.0 \times 10^{ -2 }$ & $ 7.0 \times 10^{ -2 }$  \\ 
$ 2.50 $ & $ 0.11 $ & $ 4.8 \times 10^{ -2 }$ & $ 8.7 \times 10^{ -2 }$  \\ 
$ 2.00 $ & $ 0.18 $ & $ 7.8 \times 10^{ -2 }$ & $ 8.5 \times 10^{ -2 }$  \\ 
\hline 
  \end{tabular}
 \caption{dN/dX ( $W_{1548} > 0.6$\,\AA) for our simulations.}
\label{tab:lciv_06_4_25.txt}
 \end{table}

\section*{Acknowledgements}

We thank Valentina D'Odorico for sharing her data on the \civ~column density function, Ben Oppenheimer and Ali Rahmati 
for sharing \civ~column density tables from their simulations with us, J.~X.~Prochaska for reading a draft 
and providing thoughtful comments, and Volker Springel for writing and allowing us to use the \arepo~code.
We also thank Kathy Cooksey for sharing her data on the \civ~equivalent width function, for many interesting discussions and for proof-reading a draft.
SB was supported by NASA through Einstein Postdoctoral Fellowship Award Number PF5-160133 and by a McWilliams Fellowship from Carnegie Mellon University.
LH is supported by NASA ATP Award NNX12AC67G and NSF grant AST-1312095.

\label{lastpage}
\bibliography{DLAfeedback}
\end{document}